\documentclass[fleqn,usenatbib]{mnras}

\usepackage{newtxtext,newtxmath}
\usepackage[T1]{fontenc}

\DeclareRobustCommand{\VAN}[3]{#2}
\let\VANthebibliography\thebibliography
\def\thebibliography{\DeclareRobustCommand{\VAN}[3]{##3}\VANthebibliography}

\usepackage{graphicx}	% Including figure files

\usepackage{amsmath}	% Advanced maths commands
\usepackage{amssymb}	% Extra maths symbols
\title[SMS formation by Super-competitive Accretion]
{Formation of supermassive stars and dense star clusters in metal-poor clouds exposed to strong FUV radiation}

\author[S. Chon, \& K. Omukai]{
Sunmyon Chon $^{1,2}$\thanks{E-mail: sunmyon@MPA-Garching.MPG.DE} and
Kazuyuki Omukai $^{2}$
\\
$^{1}$Max-Planck-Institut f$\ddot{u}$r Astrophysik, Karl-Schwarzschild-Str. 1, D-85741 Garching, Germany\\ 
$^{2}$Astronomical Institute, Graduate School of Science, Tohoku University, Aoba, Sendai 980-8578, Japan\\
}

\date{Accepted XXX. Received YYY; in original form ZZZ}

\pubyear{2024}

\begin{document}
\label{firstpage}
\pagerange{\pageref{firstpage}--\pageref{lastpage}}
\maketitle

\begin{abstract}
The direct collapse scenario, which predicts the formation of supermassive stars (SMSs) as precursors to supermassive black holes (SMBHs), has been explored primarily under the assumption of metal-free conditions. 
However, environments exposed to strong far-ultraviolet (FUV) radiation, which is another requirement for the direct collapse, are often chemically enriched to varying degrees. 
In this study, we perform radiation hydrodynamic simulations of star-cluster formation in clouds with finite metallicities, $Z=10^{-6}$ to $10^{-2} Z_{\odot}$, incorporating detailed thermal and chemical processes and radiative feedback from forming stars. 
Extending the simulations to approximately two million years, we demonstrate that SMSs with masses exceeding $10^4~M_\odot$ can form even in metal-enriched clouds with  $Z \lesssim 10^{-3} Z_{\odot}$. 
The accretion process in these cases, driven by "super-competitive accretion," preferentially channels gas into central massive stars in spite of small (sub-pc) scale fragmentation. 
At $Z \simeq 10^{-2} Z_{\odot}$, however, enhanced cooling leads to intense fragmentation on larger scales, resulting in the formation of dense star clusters dominated by very massive stars with $10^3 M_{\odot}$ rather than SMSs. 
These clusters resemble young massive or globular clusters observed in the distant and local universe, exhibiting compact morphologies and high stellar surface densities. 
Our findings suggest that SMS formation is viable below a metallicity threshold of approximately $10^{-3} Z_{\odot}$, significantly increasing the number density of massive seed black holes to levels sufficient to account for the ubiquitous SMBHs observed in the local universe. Moreover, above this metallicity, this scenario naturally explains the transition from SMS formation to dense stellar cluster formation.  
\end{abstract}

%%%%%%%%%%%%%%%%%%%%%%%%%%%%%%%%%%%%%%%%%%%%%%%%%%

%%%%%%%%%%%%%%%%% BODY OF PAPER %%%%%%%%%%%%%%%%%%

\section{Introduction}
Supermassive black holes (SMBHs) are now recognized as ubiquitous in galaxies across cosmic time, from the nearby to the distant universe. In the local universe, the Event Horizon Telescope has provided direct images of SMBHs, demonstrating that the space-time around them conforms to the predictions of general relativity \citep{EHT+2019, EHT+2022}. In the early universe, the James Webb Space Telescope (JWST) has recently unveiled a large population SMBHs at high-redshift \citep{Larson+2023, Kokorev+2023, Harikane2023c, Maiolino+2024}, 
providing crucial insights into their growth. 

Despite decades of intense study, the origin of SMBHs remains one of the most significant unsolved problems in astrophysics. The discovery of SMBHs in the very early universe offers a crucial clue to this mystery. Observations reveal that quasars already exist at redshifts as high as $z \sim 10$, corresponding to just $0.6~$Gyr after the Big Bang \citep{Goulding+2023, Maiolino+2024}. These findings suggest that either the initial seed black holes were relatively massive or the growth of these seeds occurred extraordinarily rapidly—or possibly both.

A common scenario for SMBH formation proposes that smaller seed black holes are first produced through certain processes, followed by growth via accretion of gas or mergers with other black holes \citep[e.g., ][]{Inayoshi+2020}. Super-Eddington accretion has been explored as a potential mechanism for rapid growth in the early universe \citep{Volonteri2005, Inayoshi2016}, but strong radiation and outflows from accretion disks present challenges to sustaining such a regime over extended periods \citep{Milos+2009, Sugimura+2018}. Alternatively, scenarios involving massive seed black holes have gained attention.

One promising pathway to forming such massive seeds is the so-called "direct collapse (DC)" scenario
\citep[e.g.][]{Bromm&Loeb2003, Regan+2009b}. In this model, supermassive stars (SMSs) with masses $\gtrsim 10^5~M_\odot$ are hypothesized to form in chemically pristine clouds exposed to intense far-ultraviolet (FUV) radiation. The FUV radiation dissociates H$_2$, which is the primary coolant in metal-free gas, thereby suppressing cooling and maintaining the gas temperature at $\sim 10^4~$K \citep{Omukai2001}. These elevated temperatures prevent fragmentation and allow the cloud to collapse monolithically \citep{Latif+2013, Inayoshi+2014, Kimura+2023}. Such high temperatures also lead to high accretion rates of $0.1$--$1~M_\odot\mathrm{yr^{-1}}$ onto the forming protostar. The rapid accretion suppresses stellar radiative feedback by inflating the stellar radius and lowering the effective temperature to $\sim 5000~$K, enabling continued accretion \citep{Omukai&Palla2003, Hosokawa+2012, Hosokawa+2013, Nandal+2023}. Consequently, SMSs can grow to $\sim 10^5$--$10^6~M_\odot$ within their lifetimes of $\sim$ Myr \citep{Umeda+2016, Woods+2017,  Haemmerle+2018}.
They subsequently collapse into black holes of comparable mass during or after their nuclear burning phases \citep{Shibata+2002, Uchida+2017, Nagele+2023, Fujibayashi2024}. 
The formation of SMSs has been supported by multi-dimensional hydrodynamic simulations, even when considering fragmentation and protostellar radiative feedback \citep[e.g.][]{Becerra+2015, Sakurai+2016, Chon+2018, Matsukoba+2021}. While circumstellar disks form due to the angular momentum of accreting matter and may fragment, this fragmentation occurs infrequently and reduces the central stellar mass only moderately, leaving the overall SMS formation scenario intact.

The predicted number density of DCBHs, estimated to range from $10^{-9}$ to $10^{-3}~\mathrm{Mpc^{-3}}$ albeit with significant uncertainties \citep[e.g.][]{Dijkstra+2008, Chon+2016, Habouzit+2016, Valiante+2016, Wise+2019, Latif+2022, Toyouchi+2023, Kiyuna+2024}, falls far short of the number density of SMBHs in the present-day universe, approximately $0.1~\mathrm{Mpc^{-3}}$ \citep{Aller+2002, Davis+2014}. The number density of DCBHs successfully explains the existence of SMBHs at very high redshifts ($z \gtrsim 7$), with a number density of $\sim 1~\mathrm{Gpc^{-3}}$ \citep{Willott+2010}. However, their properties and redshift distribution show no significant discontinuity compared to lower-redshift SMBHs, suggesting the need for a universal scenario to explain SMBH origins across all redshifts, rather than relying solely on the DCBH scenario for the highest redshifts.

The expected low number density of DCBHs stems from the stringent conditions required for their formation: the clouds must remain chemically pristine, be located in close proximity to luminous UV-emitting galaxies, and possess sufficient mass to undergo gravitational collapse. These conditions are often mutually exclusive. Massive clouds are typically metal-enriched to some extent, having undergone previous episodes of star formation. Also in cases where a nearby galaxy serves as a strong UV source, heavy elements from supernova explosions within the source galaxy may have been injected into clouds in the surrounding halo, contaminating them. Therefore, if the requirement for complete chemical pristine gas is relaxed, allowing for the presence of small amounts of metals, the number density of DCBHs could increase significantly.

The presence of heavy elements poses a challenge to the formation of DCBHs because even a small amount of metals or dust grains, at levels as low as $\gtrsim 10^{-5}~Z_\odot$, can induce rapid cooling of the collapsing cloud. This enhanced cooling triggers vigorous fragmentation, which prevents the formation of supermassive stars and instead leads to the formation of dense star clusters \citep{Omukai+2008}. This scenario has been widely anticipated in the literature \citep[e.g.][]{Volonteri2010}.

Nevertheless, this expectation had not been rigorously confirmed by numerical simulations until our previous study \citep{Chon&Omukai2020} (hereafter Paper I), which unexpectedly demonstrated that SMSs with masses exceeding $10^4~M_\odot$ can form even in metal-enriched clouds, provided the metallicity remains below $\sim 10^{-4}~Z_\odot$. In Paper I, we conducted three-dimensional hydrodynamic simulations of star formation in strongly FUV-irradiated, slightly metal-enriched clouds. Despite vigorous fragmentation triggered by dust cooling, most of the gas accretes onto the central massive star(s) via direct inflow or stellar mergers, enabling their growth into the SMS regime. While numerous protostars compete for the gas reservoir, the central massive star(s) are preferentially fed, a process we termed "super-competitive accretion," analogous to competitive accretion in present-day star clusters but on a much larger scale. At higher metallicities of $10^{-3}~Z_\odot$, however, the most massive stars remain below $10^3~M_\odot$ within the first $10^4$ years due to an order-of-magnitude decrease in the accretion rate caused by lower temperatures resulting from metal-line cooling.

However, in Paper I, thermal physics was not modeled self-consistently but instead approximated using barotropic equations of state pre-calculated from one-zone models that account for detailed thermal processes. Additionally, stellar radiative feedback, which can significantly influence the final mass of forming stars, was not included. For instance, ionizing radiation limits the maximum mass of Pop III stars by quenching accretion through the photoevaporation of circumstellar disks \citep{Hosokawa+2011}, while radiative heating of dust grains suppresses fragmentation induced by dust cooling by elevating dust and gas temperatures \citep{Chon+2024}. Furthermore, the simulations were limited to $10^4$ years, preventing an investigation into whether the massive stars could indeed grow to the true SMS regime of $\gtrsim 10^5~M_\odot$.

Another critical question is the nature of the stellar clusters that form in environments with low metallicity and strong FUV radiation fields. In Paper I, we found that when metal-line cooling triggers intense fragmentation at metallicities above a certain threshold, the collapsing cloud evolves into a dense star cluster rather than forming a single supermassive star. However, our simulations were limited in duration, leaving the long-term evolution of such systems unexplored.

Recent JWST observations have begun to reveal compact, dense star clusters in the early universe, exhibiting stellar surface densities comparable to those of young massive clusters (YMCs) or globular clusters observed in the local universe \citep[e.g.,][]{Vanzella+2023, Adamo+2024, Fujimoto+2024}. These findings raise several important questions: what physical conditions and processes give rise to such clusters in the early universe?  Could the interplay of low metallicity and strong radiation fields indeed favor their formation \citep[e.g.][]{Dekel+2023, Sugimura+2024}?
Notably, these compact clusters often exhibit enhanced nitrogen nebular emission, a feature rarely seen in typical galaxies \citep{Harikane+2024}. One plausible explanation for this enrichment involves the presence of SMSs within the cluster. Such a star, undergoing significant mass loss, could eject large quantities of nitrogen synthesized via the CNO cycle, potentially accounting for the observed nitrogen abundance \citep{Charbonnel+2023}. If dense clusters capable of harboring SMSs indeed form under these conditions, this mechanism might provide a unified explanation for both their compact structure and the puzzling nitrogen enrichment observed in these early-universe systems.

In this paper, we conduct updated simulations that incorporate key physical processes comprehensively. Specifically, we model the evolution of clouds self-consistently, solving the thermal physics with non-equilibrium chemistry for primordial gas and a simplified yet robust treatment of metal cooling. Furthermore, we include the effects of radiation feedback from the forming stars, which is crucial for understanding how stellar feedback influences fragmentation and mass accretion.
In addition to them, we extend our simulations to cover a timescale of $\sim$ Myr, the typical lifetimes of massive stars. This longer duration allows us to determine the final masses of supermassive stars under varying metallicities. It also enables us to examine the long-term evolution and properties of stellar systems, such as their structure, stellar population distributions, and the potential formation of dense star clusters, in FUV-irradiated, slightly metal-enriched environments.

The structure of this paper is as follows. In Section~\ref{sec::method}, we outline the numerical methodology employed in our simulations, including the treatment of non-equilibrium chemistry and radiative feedback. We also describe our approach of performing high-resolution, short-term runs to investigate the formation of low-mass fragments and low-resolution, long-term runs to examine the extended evolution of stellar systems and the growth of massive stars. Section~\ref{sec::results} presents the main findings, focusing on the formation and growth of SMSs and the properties of the resulting stellar clusters under varying metallicity conditions. 
Finally, in Section~\ref{sec::discussion}, we summarize our findings and discuss their implications for the formation of SMBHs, as well as their potential connections to the observed compact star clusters in the early universe.

\section{Methodology} \label{sec::method}
We perform radiation hydrodynamics simulations
for the gravitational collapse of clouds and star cluster formation within them using {\tt Gadget-3} \citep{Springel2005}, with several modifications. 

\subsection{Initial conditions and setup}
The initial conditions for our simulations are derived from a gas cloud, referred to as the {\it Spherical Cloud}, which was identified in the cosmological simulation by \citet{Chon+2016}. Two SMS-forming clouds were identified within a $20~h^{-1}$Mpc cosmological volume of this simulation, predicted to evolve into SMSs via the conventional direct collapse scenario. This scenario posits that SMSs form in metal-free clouds irradiated by strong far-ultraviolet (FUV) radiation from nearby galaxies. The {\it Spherical Cloud}, one of these two SMS-forming clouds, produces a protostar at $z=12.8$.
In the metal-free case, \citet{Chon+2018} demonstrated that an SMS exceeding $10^4~M_\odot$ forms in this cloud within the first $10^5$ years, based on simulations of the subsequent accretion phase. Using the same initial conditions as \citet{Chon+2018}, this study investigates the impact of varying metallicities on the final stellar masses.

We initialize our calculations when the central gas density reaches $10^2~\mathrm{cm^{-3}}$ and the gas temperature is approximately $8000~$K. At this stage, we assign a finite metallicity to all gas particles, exploring five cases with metallicities of [Z/H] $\equiv \log (Z/Z_\odot) = -2, -3, -4, -5$, and $-6$. At the starting point, the cooling time is longer than the free-fall time, meaning that the addition of metals does not immediately affect the thermal evolution of the gas.

Initially, the collapse of the cloud is followed in comoving coordinates, accounting for background expansion, until the first protostar forms. At this point, the simulation transitions to physical coordinates and we extract a spherical region with a radius of $10^6$ au centered on the highest-density peak, retaining the gas and dark matter particles within this region. The extracted region contains $\sim 10^6~M_\odot$ of gas, sufficient for the formation of SMSs.

\subsection{Chemical and radiative processes}
We solve a non-equilibrium chemical network for five primordial species: e$^-$, H, H$^+$, H$_2$, and H$^-$, including their associated cooling processes \citep{Chon+2021b}. In addition to the cooling mechanisms of primordial gas, we incorporate metal-line cooling from [CII] and [OI]. For simplicity, we assume that carbon and oxygen are entirely in the forms of C$^+$ and O, respectively, without explicitly solving their chemical reactions.

Dust thermal emission is also included as a cooling mechanism, which extracts internal energy from the gas through gas-dust collisions. The temperature of the dust grains is determined by solving the energy balance equation for the grains (see equation~\ref{eq::Tdust} below). The abundance ratio of heavy elements, along with the composition and size distribution of dust grains, is assumed to follow that of the Milky Way. The abundances of heavy elements and dust grains are scaled linearly with metallicity ($Z/Z_\odot$).

The star-forming regions considered in this study are exposed to strong FUV radiation fields. Specifically, we assume a radiation strength of $J_{21} = 1000$, consistent with the levels required for direct collapse in zero-metallicity clouds. The radiation spectrum is modeled as a black-body with a temperature of $10^4~$K.

To account for the self-shielding effect against H$_2$ dissociation, we adopt the method described by \citet{WolcottGreen+2011}. In this approach, the column density of H$_2$ exposed to the background radiation is approximated as that within the Jeans length: $N({\text{H}_2}) = \lambda_{\text{J}} n({\text{H}_2})$, where $\lambda_{\text{J}}$ is the Jeans length and $n({\text{H}_2})$ is the number density of H$_2$.

We incorporate radiative feedback from stars, accounting for four key processes: the ionization of H, the dissociation of H$_2$, the photo-detachment of H$^-$, and the heating of dust grains. 

To model the transfer of H-ionizing and H$_2$-dissociating photons, we use the RSPH method \citep{Susa2006, Chon+2017}.
In this method, the local optical depth and the column density are calculated for each SPH particle relative to the "up-wind" particle. The up-wind particle is selected from neighboring particles based on its proximity to the line of sight to the radiation source. The total optical depth and column density are obtained by summing the local contributions along the line of sight.

We separately calculate the optical depths of ionizing radiation ($\tau_\text{ion}$; $h\nu > 13.6~$eV) and FUV radiation ($\tau_\text{FUV}$; $13.6 > h\nu > 11.2~$eV). For $\tau_\text{FUV}$, we include the effects of dust absorption and scattering \citep{Nozawa+2008}, Thomson scattering, Rayleigh scattering by HI, and free-bound absorption by HI and H$^{-}$ \citep{Lenzuni+1991}.

The heating effect of stellar radiation on dust grains is incorporated through the energy balance equation for dust grains:
\begin{align} 
\label{eq::Tdust} 4 \sigma_\text{SB} T_\text{dust}^4 \kappa_\text{gr} \rho &= \Lambda_\mathrm{gas\rightarrow dust} + 4 \sigma_\text{SB} T_\text{rad}^4 \kappa_\text{gr} \rho + 4 \sigma_\text{SB} T_\text{CMB}^4 \kappa_\text{gr} \rho, 
\end{align}
where $\sigma_\text{SB}$ is the Stefan-Boltzmann constant, $T_\text{dust}$ is the dust temperature, $\kappa_\text{gr}$ is the Planck mean opacity, and $\Lambda_\mathrm{gas\rightarrow dust}$ represents the rate of energy transfer from gas to dust via collisions \citep{Hollenbach+1994}. The term $T_\text{CMB} = 2.73~\mathrm{K} (1+z) = 37.6~\mathrm{K}$ is the temperature of the cosmic microwave background (CMB) at redshift $z=12.7$.

To avoid solving the full radiation transfer for dust-heating radiation, we approximate the radiation temperature using the following expression \citep{Chon+2024}:
\begin{equation} 
T_\text{rad}^4 = \sum_i \frac{L_{*,i}}{16\pi \sigma_\text{SB} r_i^2}, 
\end{equation}
where $L_{*,i}$ is the luminosity of protostar $i$ and $r_i$ is the distance to protostar $i$. The Planck mean opacity at the dust temperature is used as the opacity for stellar radiation.
This approximation assumes that stellar radiation is rapidly absorbed by dust grains and subsequently reprocessed into thermal emission, modeled as a black body spectrum at the dust temperature. Despite its simplicity, this method accurately reproduces the temperature distribution around protostars, showing good agreement with results from flux-limited diffusion approaches. This consistency has been demonstrated across a range of environments, from low-metallicity cases \citep{Fukushima+2020a} to present-day conditions \citep{Yorke+1999, Kuiper+2018}.

\subsection{Protostellar models}

The radii and UV emissivities of accreting protostars are strongly influenced by their accretion histories. Rapid mass accretion delivers a substantial amount of entropy to the star, causing its stellar envelope to significantly expand or inflate \citep{Hosokawa+2012, Hosokawa+2013}. During this "supergiant protostar" phase, the stellar radius becomes orders of magnitude larger than that of ordinary main-sequence stars.

To account for these differences, we distinguish between two protostellar phases: the "normal protostar" phase and the "supergiant protostar" phase. The transition between these phases is determined based on the accretion history of the protostar, as described below.

\subsubsection{normal protostar phase}
We assume that protostars are initially in the "normal protostar" phase following their formation. During this phase, protostars first undergo an adiabatic accretion phase, where the accretion timescale is much shorter than the Kelvin-Helmholtz (KH) timescale. As the protostar evolves and the KH timescale becomes comparable to the accretion timescale, the star begins to cool and contract, gradually transitioning toward a main sequence structure through Kelvin-Helmholtz contraction \citep{Omukai&Palla2003}.

The stellar radius, luminosity, and effective temperature during this phase are determined using tabulated results, pre-calculated with a one-dimensional stellar evolution code assuming a fixed mass accretion rate \citep{Hosokawa+2009}. Stellar properties are provided as functions of the stellar mass $M_*$ and mass accretion rate $\dot{M}$ through interpolation of these pre-calculated tables:
\begin{align} 
R_* &= R_\text{MS} (M_*, \dot{M}), \\
L_* &= L_\text{MS} (M_*, \dot{M}) + L_\text{acc}, \\
   &= L_\text{MS} (M_*, \dot{M}) + f_\text{acc}\frac{G\dot{M}M_*}{R_*},
\end{align}
where $f_\text{acc}$, the fraction of accretion energy radiated away, is set to 0.75 \citep{Offner+2009}. Since the pre-calculated tables extend only up to a maximum stellar mass of $M_{,\text{max}} = 10^3~M_\odot$, we assume that for stars exceeding this mass, their luminosity equals the Eddington luminosity, and their effective temperature remains constant at the value $T_\text{eff}$ corresponding to $M = M_{*,\text{max}}$.

\subsubsection{supergiant protostar phase}
When the mass accretion rate onto a protostar is exceptionally high, the protostar enters the "supergiant protostar" phase. In this study, we adopt a critical accretion rate of $\dot{M}_\text{crit} = 0.04~M_\odot\mathrm{yr^{-1}}$ to define the transition to this phase.
Supergiant protostars are characterized by an effective temperature of $5000~$K and the Eddington luminosity. Their stellar radius is given by the following relation as a function of the stellar mass $M_*$
\citep{Hosokawa+2013}:
\begin{align} R_\text{SG} &= 2.6\times 10^3R_\odot \left ( \frac{M_*}{100M_\odot} \right )^{1/2}.
\end{align}
If the accretion rate drops below $\dot{M}_\text{crit}$, the protostar is assumed to gradually transition to the main-sequence phase. This transition occurs over a timescale corresponding to the surface KH time, which is approximated as \citep{Sakurai+2015}:
\begin{align} t_\text{KH, surf} &= 1000~\mathrm{yr} \left ( \frac{M_*}{500~M_\odot} \right )^{1/2}. \end{align}
During this transition, the stellar radius evolves according to:
\begin{align}
R_* &= \max \left \{ \frac{R_\text{SG}}{1 + \Delta t / t_\text{KH, surf}}, R_\text{MS} \right \},
\end{align}
where $\Delta t$ is the time elapsed since the accretion rate fell below $\dot{M}_\text{crit}$, and $R_\text{MS}$ is the radius corresponding to the main-sequence phase. This framework allows for a smooth and realistic transition in the stellar structure as the accretion conditions evolve.

\subsection{Particle splitting}
To resolve regions prone to gravitational instability during cloud collapse, we employ particle splitting. This technique, based on \citet{Kitsionas+2002}, divides a single gas particle into 13 daughter particles. Particle splitting is triggered when the gas density exceeds specific threshold values, $n_\text{split}$, which are set depending on the metallicity. As higher metallicity leads to lower typical temperatures and Jeans masses, correspondingly lower density thresholds for particle splitting are adopted \citep{Omukai+2008}. Specifically, for metallicities of [Z/H] $=-4$ to $-6$, the splitting thresholds are $n_\text{split} = 10, 10^3, 10^7, 10^9, 10^{11}$, and $10^{13}~\mathrm{cm^{-3}}$. For higher metallicities of [Z/H] $=-2$ and $-3$, the thresholds are set at $n_\text{split} = 10, 10^3, 10^5, 10^7, 10^9$, and $10^{11}~\mathrm{cm^{-3}}$.
Initially, the mass of an unsplit gas particle is $268~M_\odot$. With successive splitting at higher densities, the particle mass reduces to $5.56\times 10^{-5}~M_\odot$ at the highest splitting level. This level of resolution ensures that the Jeans mass is represented by more than 1000 SPH particles, meeting the criteria for accurately tracking gravitational collapse and fragmentation \citep{Bate+1995}.

\subsection{Capturing early and long-term evolution with two resolution classes}
Once the gas density exceeds $n_\text{adia}$, we assume the gas evolves adiabatically, neglecting any further cooling processes. Sink particles are introduced when the gas density surpasses $n_\text{sink}$ and the particle is located at the local minimum of the gravitational potential \citep{Hubber+2013}. To prevent spurious sink formation, we set $n_\text{sink} = 2 \times n_\text{adia}$, as adopted in previous studies \citep{Chon+2018, Susa2019}.

Our simulations consist of two classes with distinct numerical resolutions. The high-resolution, short-term runs employ $n_\text{sink} = 2 \times 10^{16}~\mathrm{cm^{-3}}$, allowing detailed tracking of the early phases of protostar formation. These runs use $n_\text{adiabatic} = 10^{16}~\mathrm{cm^{-3}}$, corresponding to the opacity limit typically associated with SMSs. This configuration ensures the protostellar radii of massive stars are well-resolved, with each sink particle representing an individual protostar. These runs cover the initial evolution for up to $10^4$ years for [Z/H]$\lesssim -4$, $10^5$ years for [Z/H]$=-3$, and $3 \times 10^5$ years for [Z/H]$=-2$, following the formation of the first protostar. At the end of the simulations, the protostars are still actively accreting and increasing in mass.
To capture the long-term evolution of the system, we also perform low-resolution, long-term runs with $n_\text{sink} = 2 \times 10^{11}~\mathrm{cm^{-3}}$. These simulations extend over 2 Myr after the formation of the first protostar, sufficient to track the evolution until the mass accretion ceases. While the lower resolution does not resolve the formation of low-mass stars with $\lesssim 1~M_\odot$, it successfully reproduces the distribution of massive stars (see Appendix~\ref{sec::resolution}).

By combining these approaches, we analyze the initial fragmentation and formation processes in detail with the high-resolution runs and examine the long-term evolution and mass distributions of stellar systems with the low-resolution runs. Additionally, the missing low-mass stellar population in the low-resolution runs can be inferred from the results of the high-resolution simulations. This integrated approach enables a comprehensive investigation of both the early and long-term evolution of protostellar systems and the conditions necessary for the formation of SMSs.

\begin{figure}
\centering
\includegraphics[width=0.45\textwidth]{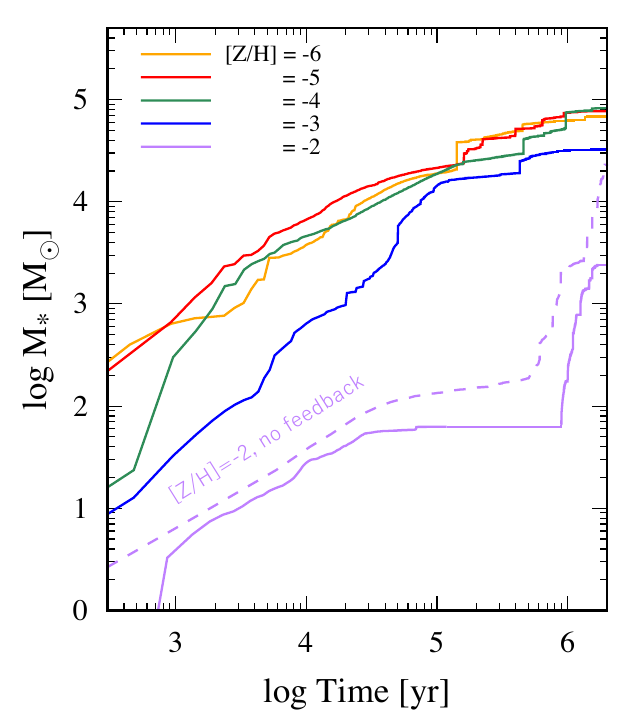}
\caption{
Growth histories of the most massive stars at the end of the simulations for different metallicities: [Z/H]$=-2$ (purple), [Z/H]$=-3$ (blue), [Z/H]$=-4$ (green), [Z/H]$=-5$ (red), and [Z/H]$=-6$ (yellow). The data are derived from the long-term, low-resolution runs. For [Z/H]$=-2$, the dashed line represents the case without stellar feedback, while for the other metallicities, the results with and without feedback are similar, and thus only the cases with feedback are shown. In the very low metallicity cases of [Z/H]$\lesssim -3$, the most massive stars grow to the SMS regime, reaching masses of $\gtrsim 3 \times 10^4 M_{\odot}$. In contrast, at [Z/H]$=-2$, the stellar mass is significantly smaller, barely exceeding $2000~M_\odot$, which is more than an order of magnitude lower than in the lower metallicity cases.
}
\label{fig::mass_evolution}
\end{figure}

\begin{figure*}
\centering
\includegraphics[width=0.95\textwidth]{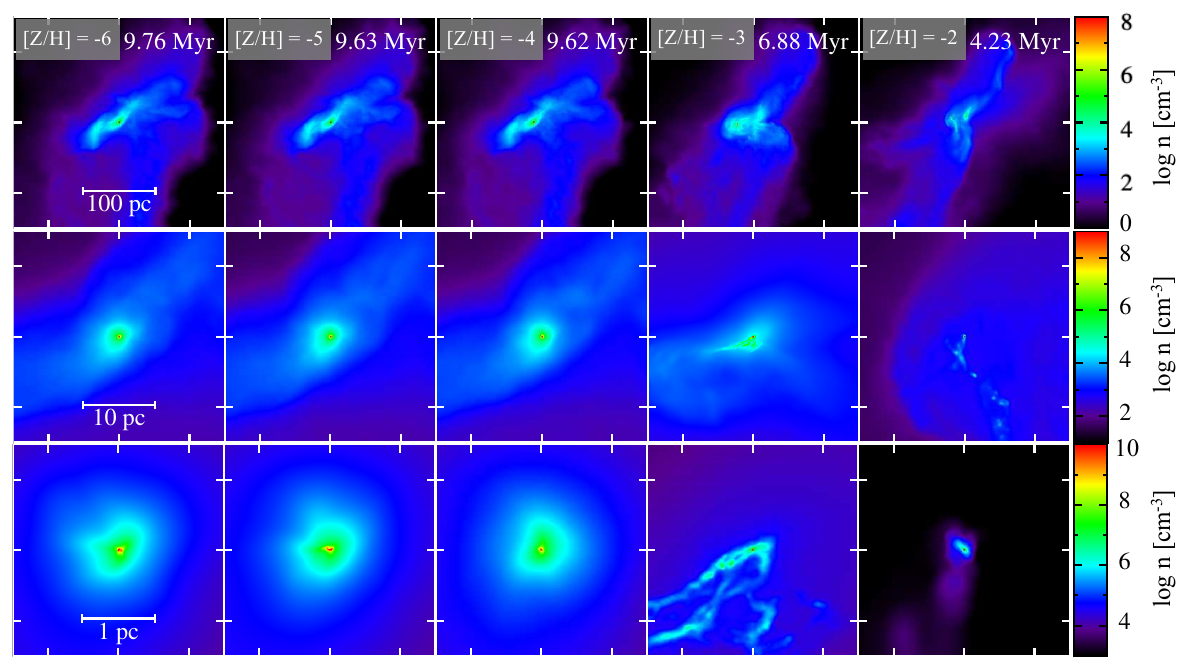}
\caption{
Projected density distributions at three different spatial scales for various metallicities, [Z/H]$=-6$ to $-2$ (from left to right), taken at the epoch when the first protostar forms. The snapshots are from the high-resolution runs, which resolve gas densities up to $\sim 10^{16}~\mathrm{cm^{-3}}$. Each column corresponds to a different metallicity and shows the distribution at progressively smaller scales. Note that each panel uses a different color scale, corresponding to the varying dynamic ranges of density across the different spatial scales. The times elapsed since the onset of the calculations are shown in the top-left corners of the top panels. In the left three cases with [Z/H] $\leq -4$, the density structures are similar across all three scales presented. In contrast, the right two cases with [Z/H] $\geq -3$ exhibit significant differences from the lower metallicity cases, particularly at the smaller scales shown in the bottom two rows.
}
\label{fig::snap_overall_density}
\end{figure*}

\begin{figure*}
\centering
\includegraphics[width=0.95\textwidth]{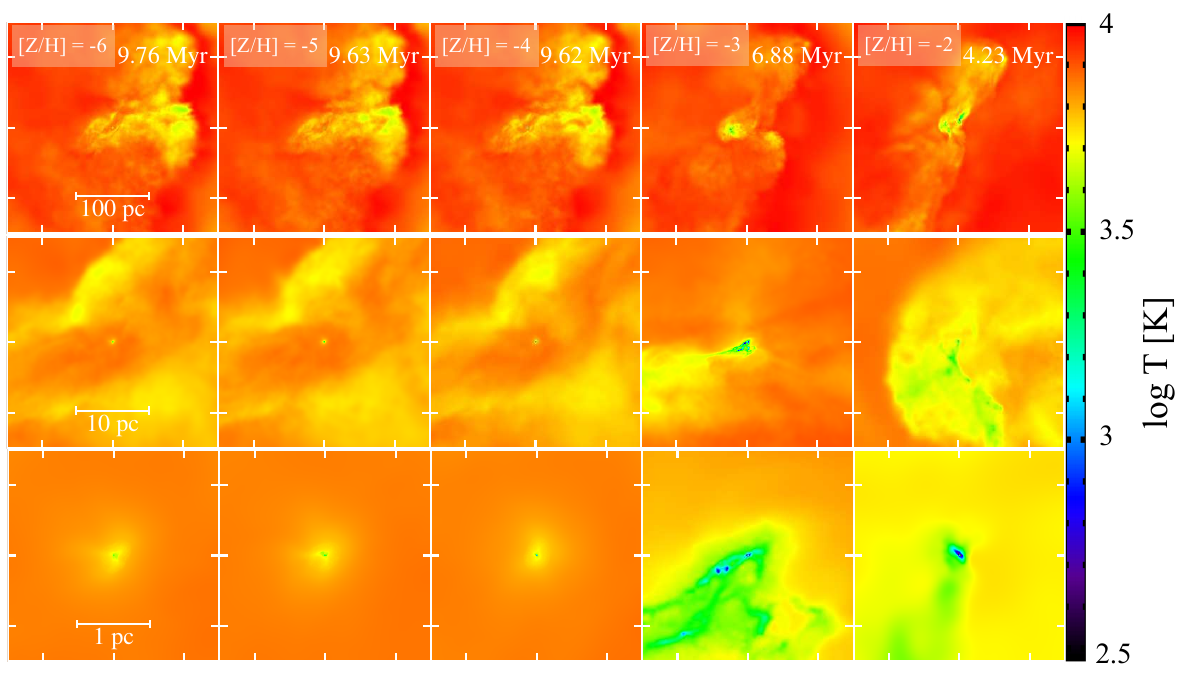}
\caption{
Temperature distributions corresponding to the density distributions shown in Fig.~\ref{fig::snap_overall_density}.
}
\label{fig::snap_overall_Tgas}
\end{figure*}

\begin{figure*}
\centering
\includegraphics[width=0.95\textwidth]{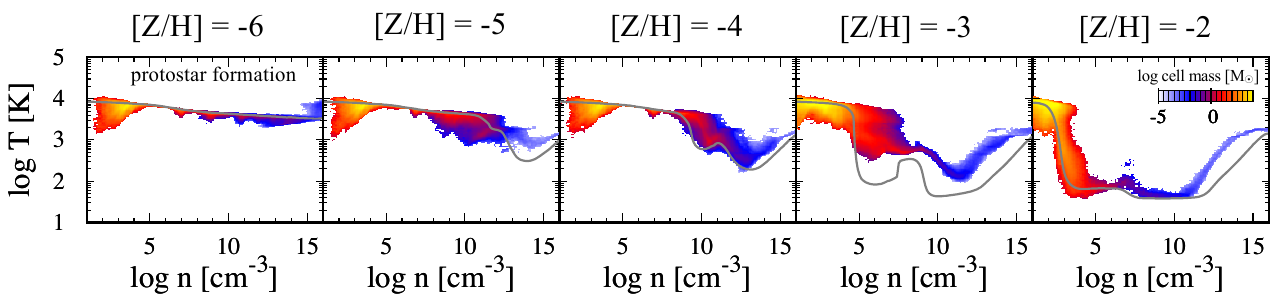}
\caption{
The gas distributions on density-temperature phase plots at the epoch of the first protostar formation for different metallicities, [Z/H]$=-6$, $-5$, $-4$, $-3$, and $-2$, from left to right panels. The plots are derived from high-resolution runs. The color coding represents the gas mass within each grid, with the density-temperature diagram divided into $200 \times 200$ grids. Gray lines show the temperature evolution predicted by one-zone models, which assume that the density increases at the free-fall rate \citep[e.g.][]{Omukai+2008}. Our numerical simulation results generally agree well with the one-zone model predictions, although there is notable deviation in the [Z/H]$=-3$ case.
}
\label{fig::rhoT_hist}
\end{figure*}

\section{Result} \label{sec::results}
The collapse of the cloud in each simulation results in the formation of a star cluster. The properties of these star clusters, including the mass of the most massive star, are strongly influenced by the metallicity of the cloud.

In Section~\ref{sec::overall}, we provide an overview of the collapse and large-scale fragmentation of star-forming clouds, with a particular focus on the formation and growth of central massive stars as a function of metallicity. Next, in Section~\ref{sec::early_development}, we explore the early development of star clusters, highlighting the processes of small-scale fragmentation and the subsequent growth of central massive stars within these clusters. Finally, in Section~\ref{sec::formation_of_stellar_system}, we focus on the detailed properties of the stellar systems that emerge from these processes.

\subsection{Cloud collapse, large-scale fragmentation and formation of the central most massive stars} \label{sec::overall}

We begin by discussing the mass of the most massive star at the end of each simulation. Fig.~\ref{fig::mass_evolution} shows the growth histories of these stars for different metallicity cases. These results are derived from the long-term, low-resolution runs, which track the mass evolution over 2 Myr, the typical lifetime of massive stars. It is important to note that while the low-resolution runs may affect the number and masses of low-mass stars, the properties of massive stars remain unaffected (see Section~\ref{sec::method} and Appendix~\ref{sec::resolution}). 

For very low metallicities of [Z/H]$\lesssim -4$, the stellar masses grow steadily over time at a rate of approximately $1~M_\odot\mathrm{yr^{-1}}$. Eventually, they reach the SMS regime, with final masses of $6$--$8 \times 10^4~M_\odot$, consistent with the conventional direct-collapse scenario under pristine gas conditions and in agreement with the findings of Paper I.
At a slightly higher metallicity of [Z/H] = -3, the stellar mass remains an order of magnitude smaller than in the lower-metallicity cases during the initial $\sim 10^4$ years. However, it eventually catches up by $t \gtrsim 10^5$ years, reaching a final mass in the SMS regime of $3 \times 10^4~M_\odot$. While slightly lower than the masses achieved in the lower-metallicity cases, it still represents significant growth.
It is worth noting that in Paper I, this late-time efficient growth observed in the [Z/H] = -3 case was not captured, as the simulations were limited to the initial evolution up to $10^4$ years.
At the highest metallicity considered, [Z/H] = -2, the most massive star remains below 100 $M_{\odot}$ during the first 1 Myr of evolution. Subsequently, it undergoes a phase of rapid mass growth driven by the influx of substantial gas, as discussed later. However, despite this late-stage acceleration, the star ultimately reaches only a few thousand $M_{\odot}$, falling short of the SMS mass range. This limited growth is likely a consequence of the reduced accretion efficiency caused by fragmentation at larger spatial scales and the effects of stellar feedback (see later), which inhibit the sustained accumulation of mass required to form an SMS.

\begin{figure}
\centering
\includegraphics[width=0.4\textwidth]{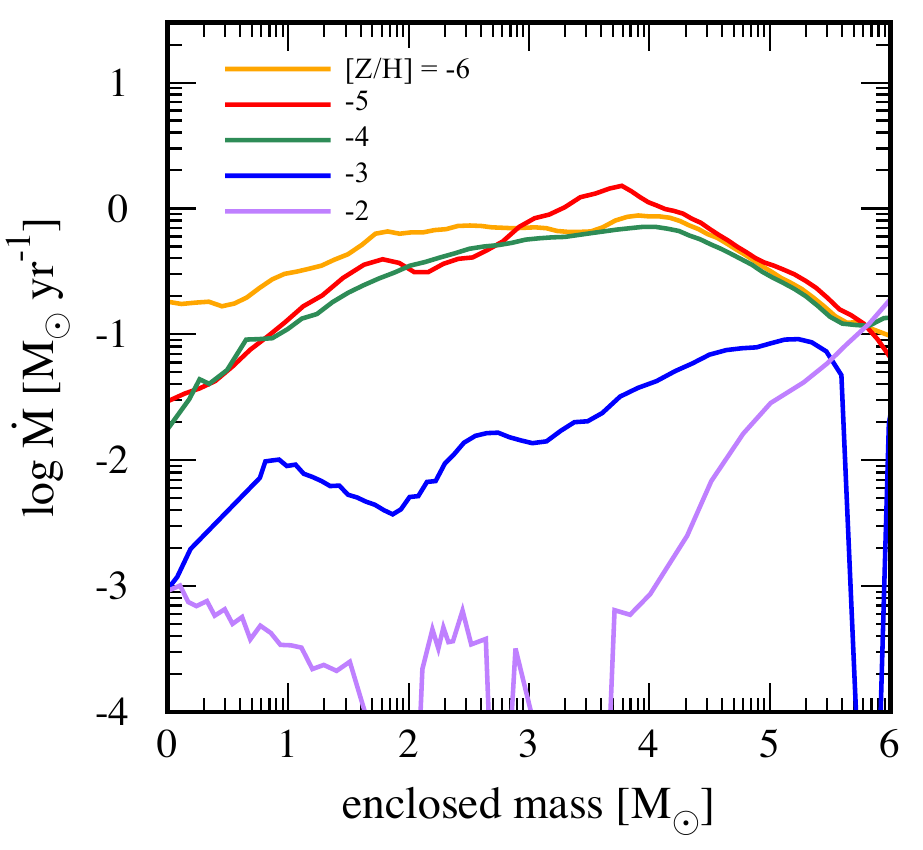}
\caption{
The gas infall rate, estimated from the snapshot at the time of the first protostar formation, as a function of the enclosed mass within a given radius (see text) for different metallicities: [Z/H]$=-2$ (purple), [Z/H]$=-3$ (blue), [Z/H]$=-4$ (green), [Z/H]$=-5$ (red), and [Z/H]$=-6$ (yellow lines). This analysis is based on the short-term, high-resolution runs, which effectively capture the early phase of evolution at the time of the first protostar formation.
}
\label{fig::mdot}
\end{figure}

The results above indicate that SMSs successfully form for [Z/H]$\lesssim -3$ but fail for [Z/H]$=-2$. This outcome is primarily due to differences in cloud morphology and the resulting fragmentation on relatively large scales of $\gtrsim 1$--$10$ pc. The overall cloud structures at the time of the first protostar formation are illustrated in Figs.~\ref{fig::snap_overall_density} and \ref{fig::snap_overall_Tgas}, which show the density and temperature distributions, respectively, at different scales for various metallicities.
We also present the thermal evolution of the clouds in Fig.~\ref{fig::rhoT_hist}, which depicts gas mass distributions on temperature-density diagrams at specific epochs. These snapshots are taken from the short-term, high-resolution runs, resolving high-density regions up to $\sim 10^{16}~\mathrm{cm^{-3}}$.
The top rows in Figs.~\ref{fig::snap_overall_density} and \ref{fig::snap_overall_Tgas} show the distributions on scales of approximately 100 pc.
The cloud structure at scales larger than $100~$pc is similar in all cases: a long filamentary structure with a density of $\sim 100~\mathrm{cm^{-3}}$ extends vertically, while the temperature consistently remains around $\gtrsim 6000$ K in most of the regions. 
As shown in the density-temperature diagrams (Fig.~\ref{fig::rhoT_hist}), low-density gas ($\lesssim 100~\mathrm{cm^{-3}}$) maintains a temperature of several 1000 K, irrespective of metallicity, due to the inefficiency of metal-line cooling at these densities.
Consequently, the density structure in these low-density regions shows little dependence on metallicity.
In contrast, as evident in the middle and bottom rows of Fig.~\ref{fig::snap_overall_density}, showing the density distributions at smaller scales, the structure of the central cloud cores, spanning a few tens of parsecs, exhibits significant dependence on metallicity.  This structural variation arises from differences in temperature evolution at higher densities ($\gtrsim 10^3~\mathrm{cm^{-3}}$) across different metallicities, as we will discuss shortly.

In the very low-metallicity cases of [Z/H]$\lesssim -4$, a highly concentrated region of dense ($\sim 10^6~\mathrm{cm^{-3}}$) and warm ($\sim 6000~$K) gas, spanning several pc, is observed near the center (middle row in the three left columns of Figs.~\ref{fig::snap_overall_density} and \ref{fig::snap_overall_Tgas}). 
Within this dense region, a spherical core of approximately 1 pc is observed, with no indications of fragmentation at this scale (bottom-row panels of Figs.~\ref{fig::snap_overall_density} and \ref{fig::snap_overall_Tgas}). This is because, in these cases, the gas remains nearly isothermal at several thousand Kelvin, without undergoing a rapid cooling phase at densities below $\sim 10^{10}~\mathrm{cm^{-3}}$, thereby preventing fragmentation (Fig.~\ref{fig::rhoT_hist}).
At even higher densities, the temperature evolution differs among these cases. For the lowest metallicity, [Z/H]$=-6$, no significant temperature drop is observed, similar to the behavior of primordial gas. In contrast, for [Z/H]$=-5$ and $-4$, dust cooling induces a rapid temperature decline at densities above $\sim 10^{10}~\mathrm{cm^{-3}}$, resulting in small-scale fragmentation on sub-parsec scales, while not affecting the overall cloud morphology on scales greater than 0.1 pc shown in Fig.~\ref{fig::snap_overall_density}. This temperature decrease also slightly accelerates the collapse, as indicated by the collapse times: 9.76 Myr for [Z/H]$=-6$ compared to 9.62 Myr for [Z/H]$=-4$.

In contrast, at higher metallicities [Z/H]$\gtrsim -3$, clouds with $\gtrsim 10^4~\mathrm{cm^{-3}}$ undergo fragmentation into smaller cores (see panels in the two right columns of the middle and bottom rows in Figs. \ref{fig::snap_overall_density} and \ref{fig::snap_overall_Tgas}). For [Z/H]$=-3$, a filament with a density of $\sim 10^6~\mathrm{cm^{-3}}$ and a temperature of $\sim 10^3$ K, extending over a $\sim$ pc scale, forms (Figs.~\ref{fig::snap_overall_density} and \ref{fig::snap_overall_Tgas}) due to a temperature drop to $\sim 10^3~$K at $\sim 10^{5}~\mathrm{cm^{-3}}$ caused by metal-line cooling (Fig.~\ref{fig::rhoT_hist}). This filament subsequently fragments into smaller cloud cores with typical separations of $\sim$ pc (Fig. \ref{fig::snap_overall_density}), driven by further cooling to $\sim 100~$K via dust cooling at densities of $ \sim 10^{9}~\mathrm{cm^{-3}}$ (Fig. \ref{fig::snap_overall_Tgas}).
However, due to their small separations, all these fragments eventually fall toward the cloud center and merge with the central core. For [Z/H]$=-2$, a filamentary structure forms in a more diffuse ($\sim 10^4~\mathrm{cm^{-3}}$) and extended region of $\sim 10$ pc compared to [Z/H]$=-3$ (Fig. \ref{fig::snap_overall_density}, middle panel), as a result of a sudden temperature drop shortly after the onset of collapse at $\sim 10^2$--$10^3~\mathrm{cm^{-3}}$
from several 1000 K to approximately 100 K, driven by metal-line cooling (Fig.~\ref{fig::rhoT_hist}).
The filament maintains a temperature of $\sim 100$ K and is surrounded by warmer, lower-density gas (Fig.~\ref{fig::snap_overall_density}). It immediately fragments into several cores, separated by distances of 1--10 pc, which are more isolated compared to the case of [Z/H]$=-3$. This larger separation arises from the earlier onset of the temperature drop, occurring at lower densities due to more efficient cooling.
At higher densities, the temperature remains nearly constant at several 10 K over many orders of magnitude in density $10^{3}-10^{11} {\rm cm^{-3}}$ (Fig.~\ref{fig::rhoT_hist}), even though dust becomes the dominant coolant at $\sim 10^{6}~\mathrm{cm^{-3}}$, instead of metal-line cooling. At densities around $\sim 10^{11}~\mathrm{cm^{-3}}$, the cloud becomes optically thick to dust attenuation, causing the temperature to increase adiabatically thereafter.

It is worth noting that the temperature evolution of the collapsing clouds is similar to that predicted by one-zone calculations, which follow the evolution of the cloud central region by assuming that the density increases at the free-fall rate, as described in \citet{Omukai+2008}. The gray lines in Fig.~\ref{fig::rhoT_hist} represent the temperature evolution predicted by the one-zone model, which closely matches the simulation results, except in the case of [Z/H]$=-3$. This similarity indicates that the collapse rate is generally consistent with the free-fall rate assumed in the one-zone model. The discrepancy at [Z/H]$=-3$ arises because the cloud collapses more rapidly than the free-fall time, driven by higher pressure in the outer regions. This results in enhanced adiabatic heating and, consequently, higher temperatures compared to the one-zone model prediction.

In the densest central region of the cloud, a protostar forms and rapidly grows in mass through the accretion of surrounding gas, eventually becoming very massive.
The mass growth of the most massive stars, as shown in Fig.~\ref{fig::mass_evolution}, can be explained by the mass infall rate at the time of the first protostar formation. This is illustrated in Fig.~\ref{fig::mdot}, which presents the infall rate as a function of the enclosed mass, $M_{r}$, within a given radius, $r$, from the cloud center for each metallicity.
The infall rate, $\dot{M}_{\rm in}$, is calculated using:
\begin{align}
\dot{M}_{\rm in} = 4 \pi r^2 \rho (r) v_\text{in} (r),
\end{align}
where $\rho(r)$ and $v_\text{in}(r)$ are the gas density and radial velocity of the gas at radius $r$ from the center at the time of 
the first protostar formation.
The infall rate at a given enclosed mass shown in Fig.~\ref{fig::mdot} approximates the accretion rate of the star as it grows and reaches the corresponding enclosed mass value.

Before examining the distribution of infall rates for different metallicities, we overview how these rates are linked to the thermal evolution of the collapsing gas. When a protostar forms, the density structure of the surrounding envelope is approximately that of a singular isothermal sphere, described by $\rho \sim c_{\rm s}^2 / 2 \pi G r^2$, with a modest enhancement factor of order unity. The infall velocity is roughly equal to the sound speed, $c_{\rm s}$. As a result, the infall rate can be expressed as \citep{Larson1969, Shu1977, Whitworth&Summers1985}:
\begin{equation}
    \dot{M} \simeq \phi c_{\rm s}^3/G \propto T^{3/2}
\end{equation}
where $\phi$ is a non-dimensional factor that accounts for the dynamical nature of the collapse. This includes factors such as whether the gas was initially at rest or dynamically collapsing, as well as the influence of turbulence, or rotation.
The equation highlights that regions with higher temperatures lead to higher infall rates, and thus the higher mass growth rates of the protostar.

When the metallicity is very low at [Z/H]$\lesssim -4$, the mass infall rate consistently evolves across all cases, peaking at approximately $1~M_\odot \mathrm{yr}^{-1}$ around an enclosed mass of $M_r = 10^4~M_\odot$. This high accretion rate is driven by the high temperature of the collapsing gas, nearly constant at $\sim 10^4~$K. This peak corresponds to a gas density of $10^{6}~\mathrm{cm^{-3}}$, where the temperature is maintained at $\sim 10^4~$K across this metallicity range (Fig.~\ref{fig::rhoT_hist}).
Beyond this radius, the infall rate decreases to $0.1~M_\odot \mathrm{yr}^{-1}$ at an enclosed mass of $M_r = 10^6~M_\odot$. This decline aligns with a reduction in external density, likely caused by truncation due to an external tidal field \citep{Chon+2016}. For these low-metallicity cases, the mass growth of the central star slows significantly after reaching $\sim 10^4~M_\odot$ at $\gtrsim 10^5$ years, as shown in Fig.~\ref{fig::mass_evolution}, corresponding to the reduction in $\dot{M}$ beyond $M_r = 10^4~M_\odot$.

At a slightly higher metallicity of [Z/H]$=-3$, the infall rate in the inner regions is about an order of magnitude lower than that in the lower-metallicity cases but increases toward the outer regions. Specifically, the infall rate is $\lesssim 10^{-2}~M_\odot \mathrm{yr}^{-1}$ within $M_r = 10^3~M_\odot$, rising to approximately $0.1~M_\odot \mathrm{yr}^{-1}$ around $10^4$--$10^5~M_\odot$.
This outward increase in the infall rate can be attributed to the temperature structure within the cloud. In the central, high-density region ($\gtrsim 10^6~\mathrm{cm^{-3}}$), metal-line cooling reduces the temperature to approximately 500 K. In contrast, the temperature remains at $\sim 10^4~$K in the outer, lower-density regions due to the inefficiency of cooling processes at these densities (see Fig.~\ref{fig::rhoT_hist}). Consequently, the inner, cooler regions exhibit a lower accretion rate of $10^{-2}~M_\odot \mathrm{yr^{-1}}$, while the hotter outer regions support a more substantial inflow of $\sim 0.1~M_\odot \mathrm{yr^{-1}}$.
The hot inflowing gas reaches the central region within approximately the free-fall time of $\sim 10^5$ years, resulting in an increase in the mass accretion rate during this period. While fragmentation occurs on scales of 0.1 to 1 pc in this metallicity case (Fig.~\ref{fig::snap_overall_density}), the small cloud cores formed by fragmentation eventually migrate through dense filaments and accrete onto the central star. As a result, the accretion flow remains concentrated on the central star rather than being divided among multiple stars. This concentrated growth is evidenced by the mass evolution shown in Fig.~\ref{fig::mass_evolution}, which aligns closely with the infall rate trends illustrated in Fig.~\ref{fig::mdot}.

For [Z/H]$=-2$, the highest metallicity considered, the mass inflow rate remains relatively low at $\lesssim 10^{-3}~M_\odot \mathrm{yr^{-1}}$ within the cold inner regions, primarily due to efficient metal-line cooling, which significantly reduces the gas temperature. In contrast, in the outer regions, where the enclosed mass exceeds $10^5~M_\odot$, the inflow rate can reach values as high as $10^{-2}$ to $10^{-1}~M_\odot \mathrm{yr^{-1}}$, supported by the high temperature of $\sim 10^4~$K in these regions.
However, this substantial inflow undergoes fragmentation into numerous low-mass stars before reaching the central region. This fragmentation is driven by rapid metal-line cooling, which becomes effective at densities of $\sim 10^3~\mathrm{cm^{-3}}$. The fragments, although they possess inward radial velocities and migrate toward the central protostar, fail to reach the central region within the stellar lifetime due to the wide spatial scale of fragmentation.
As a result, the inflowing gas is divided among multiple fragmented cores, preventing efficient accretion onto the central star(s). Consequently, the formation of a SMS is not achieved in this case.

In these simulations, unlike in Paper I, we incorporated radiative feedback from the forming protostars, enabling us to investigate its impact on the final stellar mass. However, for very low metallicities of [Z/H]$\lesssim -3$, radiative feedback had little effect on the mass growth of SMSs. In contrast, it significantly reduced the stellar mass in the case of [Z/H]$=-2$.
The minimal impact of radiative feedback in the very low-metallicity cases can be attributed to the high accretion rates, which result in extremely dense regions surrounding the protostars. These dense environments efficiently attenuate ionizing photons, preventing the ionized regions from expanding significantly. As a result, photo-heating has a negligible influence on the accretion flow and the mass evolution for [Z/H]$\lesssim -3$ remains nearly identical between cases with and without radiative feedback. For this reason, the mass evolution without feedback is not explicitly shown in Fig.~\ref{fig::mass_evolution}.
\begin{figure*}
\centering
\includegraphics[width=1.\textwidth]{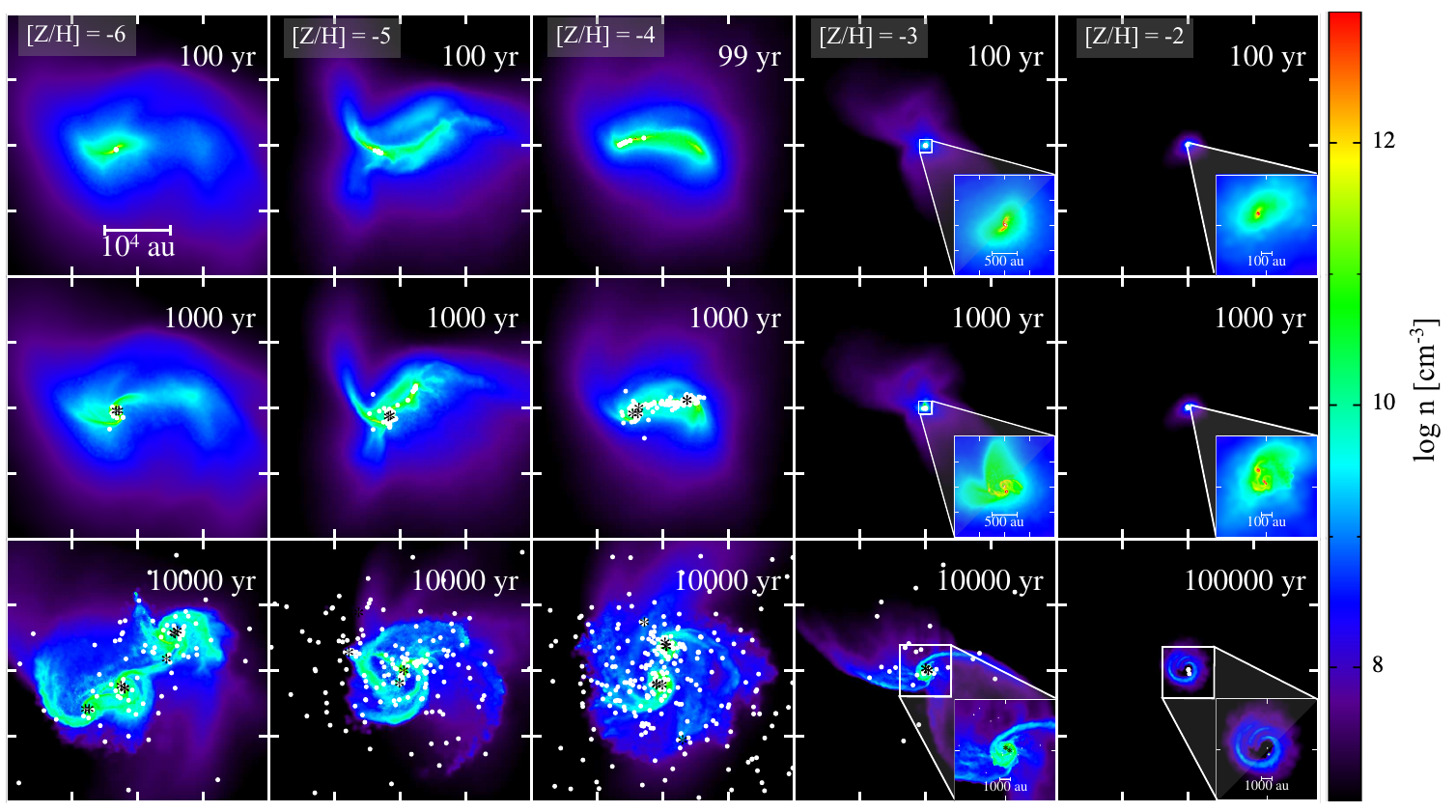}
\caption{
Projected density distributions at three different epochs following the formation of the first protostar for various metallicities: [Z/H]$=-6$, $-5$, $-4$, $-3$, and $-2$, from left to right. The top and middle rows show the distributions at 100 years and 1,000 years, respectively, while the bottom row presents the distributions at 100,000 years for [Z/H]$=-2$ and at 10,000 years for [Z/H]$\lesssim -3$, after the formation of the first protostar. Asterisks indicate the positions of protostars with masses greater than $100~M_\odot$, while dots mark those with masses less than $100~M_\odot$
}
\label{fig::snap_evo}
\end{figure*}
For [Z/H]$=-2$, however, fragmentation occurs on much larger spatial scales compared to lower-metallicity cases (Fig.~\ref{fig::snap_overall_density}). This results in lower densities around the protostars, allowing ionized regions to expand more readily. Stellar radiative feedback heats and evaporates the surrounding gas, ultimately suppressing the formation of SMSs.
The impact of radiative feedback on SMS formation is clearly demonstrated by comparing the purple solid and dashed lines in Fig.~\ref{fig::mass_evolution}. In the absence of stellar feedback (dashed line), the stellar mass exceeds $10^4~M_\odot$. In contrast, when feedback is included, the stellar mass is reduced by an order of magnitude, barely reaching $10^3~M_\odot$. Although the growth of SMSs is significantly hindered, star formation within the cloud persists until the end of our simulation, which extends to 2 Myr after the formation of the first protostar.

\subsection{Early development of star clusters and growth of very/super-massive stars within them}
\label{sec::early_development}
Sub-pc-scale fragmentation, occurring at distances less than $0.05~$pc (i.e., $10^4$ au), is observed across all metallicities in our simulations. Unlike the larger-scale fragmentation observed at 1--10 pc in the case of [Z/H]$=-2$, which inhibits the central stars from achieving very high masses, this small-scale fragmentation does not impede the formation of SMSs for metallicities of [Z/H]$\lesssim -3$. This suggests that the impact of fragmentation on SMS growth is scale-dependent, with sub-pc fragmentation having a minimal effect on the accretion processes that drive the formation of these massive stars.
Small-scale fragmentation results in the formation of three to six SMSs, which collectively establish a hierarchical multiple system at the center of the cloud. This process also gives rise to numerous low-mass stars. However, despite the occurrence of sub-pc-scale fragmentation, the majority of the mass consistently accretes onto the central massive stars, indicating that the growth of these stars remains largely unaffected by the presence of smaller fragments.

The density structures of the cloud at three different epochs—100 years, 1000 years, and 10,000 years after the formation of the first protostar (or 100,000 years for [Z/H]$=-2$)—are shown in Fig.~\ref{fig::snap_evo}. These snapshots are derived from the short-term, high-resolution runs.
As illustrated, small-scale fragmentation occurs differently across these epochs depending on the metallicity of the cloud. However, as will be discussed later, this small-scale fragmentation has negligible impact on the mass evolution of the central massive stars. In the following sections, we examine the metallicity-dependent nature of this fragmentation and its implications for star formation.

\begin{figure}
\centering
\includegraphics[width=0.45\textwidth]{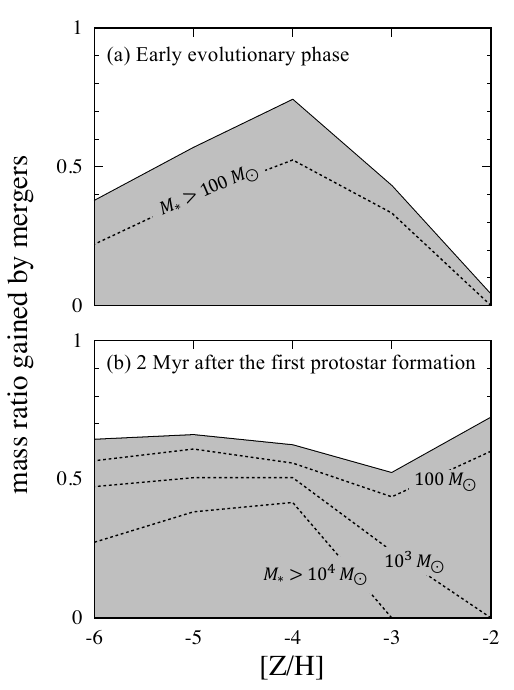}
\caption{
Fraction of the mass of the most massive star in each simulation that was acquired through stellar mergers, as opposed to gas accretion, at the end of the respective runs. The top panel shows results from the short-term, high-resolution runs at $10^4$ years for all metallicities except [Z/H]$=-2$, where it is $10^5$ years. The bottom panel shows results from the long-term, low-resolution runs at the end of 2 Myrs.
The dashed lines represent the contributions from mergers with stars of $M_* > 100$, $10^3$, and $10^4~M_\odot$, from top to bottom, respectively.
}
\label{fig::mass_by_mergers}
\end{figure}

We begin by examining the lowest metallicity case, [Z/H]$=-6$. In this case, dust cooling is negligibly effective, making the conditions nearly identical to those of a metal-free cloud. As a result, the conventional direct-collapse pathway for SMS formation is realized. The cloud undergoes monolithic collapse without significant fragmentation, giving rise to a single massive protostar at its center.
Around this central protostar, a circumstellar disk forms. By $t=10^3~$yr, fragmentation within the disk occurs, leading to the formation of several low-mass stars confined within a radius of approximately 1000 au (middle panel).
Some of the stars formed through disk fragmentation grow significantly, reaching masses of $10^3$--$10^4~M_\odot$. By $t=10^4$ years (bottom panel), multiple rotationally supported disks develop around these massive stars, driven by the finite angular momentum of the accreting gas.
These disks continue to grow in mass by accreting gas from the surrounding envelope. However, gravitational instability within the disks triggers further fragmentation, ultimately leading to the formation of six very massive or supermassive stars with $M_* \gtrsim 1000~M_\odot$. They establish a hierarchical multiple system at the center of the cloud.
Lower-mass stars ($<100~M_\odot$) also form through disk fragmentation, facilitated by H$_2$ cooling, alongside the massive stars. However, the orbits of these low-mass stars are dynamically unstable due to the strong gravitational influence of the central massive stars, leading to their ejection from the natal disk.
Although numerous protostars are formed over time, a few central protostars dominate the accretion of disk gas. Their strong gravitational pull enables them to accrete gas almost exclusively, allowing them to grow to supermassive, ultimately reaching masses of $M_* \gtrsim 10^4~M_\odot$ by $t=10^4$ years.

\begin{figure*}
\centering
\includegraphics[width=1.\textwidth]{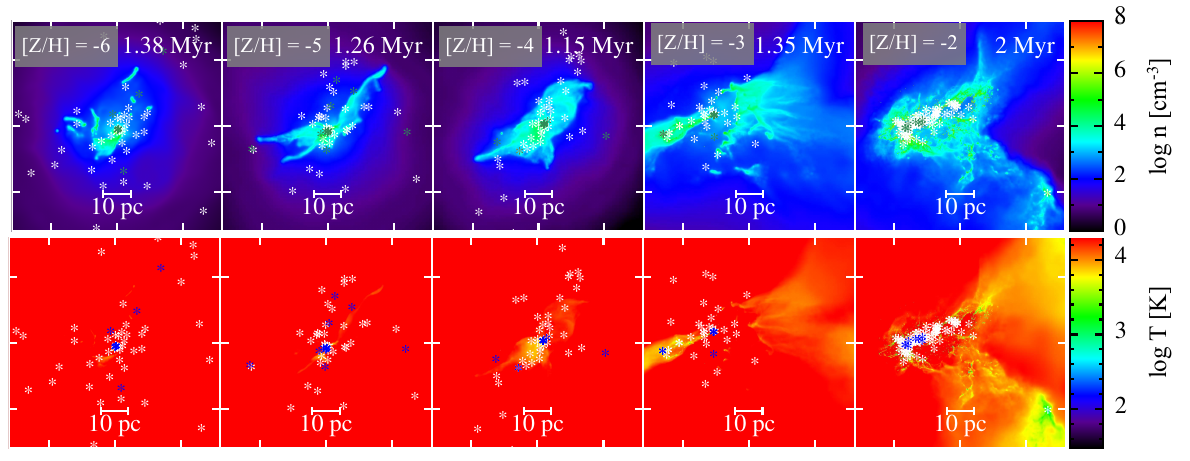}
\caption{
Cloud structures for different metallicity cases ([Z/H]=-6,..., -2 from left to right columns) 
at the onset of cloud evaporation due to stellar radiative feedback. The top panels display the projected density distributions, while the bottom panels show the corresponding temperature distributions. The times since the formation of the first protostar are indicated in the top-right corners of the top panels. White asterisks mark the locations of protostars with masses between $100$ and $1000~M_\odot$ and green/blue asterisks show those with the mass greater than $1000~M_\odot$, while lower-mass protostars are not shown.
}
\label{fig::snap_overall_tMyr}
\end{figure*}

Next, we examine the cases with slightly higher metallicities of [Z/H]$= -5$ and $-4$. At relatively low densities ($\lesssim 10^{10}~\mathrm{cm^{-3}}$), the evolution is similar to that of the [Z/H]$=-6$ case, as metallicity has little impact in this regime. However, at higher densities, enhanced dust cooling becomes significant, reducing the temperature of the collapsing gas. This effect makes the cloud more elongated, as seen at $t=100$ years in the top panel.
In the central region of the cloud, a few massive stars form at density peaks, while the filament itself fragments, producing multiple protostars distributed spatially along it over scales of $10^4$ au by $t=10^3$ years (middle panel). With increasing metallicity, dust cooling becomes more effective, leading to more vigorous fragmentation and an increase in the number of low-mass stars.
By $t=10^4$ years, the massive stars form a multiple system surrounded by circumstellar disks, similar to the [Z/H]$=-6$ case. The low-mass stars, initially formed through fragmentation along the filament, are temporarily captured within the disk region. 
Within these disks, low-mass stars with typical masses of $\sim1~M_\odot$, significantly smaller than those formed in the [Z/H]$=-6$ case, continue to form, driven by dust cooling. However, these low-mass stars are either ejected from the disk or merge with the central massive stars within a few dynamical timescales shortly after their formation.
Similarly, the disk gas is predominantly accreted by the central massive stars, whose strong gravitational influence dominates the disk dynamics. This ensures that the formation of low-mass stars does not significantly impede the growth of the central massive stellar system. By the time $t = 10^4$ years, the central stars have grown to supermassive ($M_* \gtrsim 10^4~M_\odot$), similar to what is observed in the [Z/H]$=-6$ case, while being accompanied by numerous low-mass stars. 
As proposed in Paper I, this process aligns with the concept of super-competitive accretion, where the central massive stars preferentially accrete the majority of the available disk gas. This mechanism enables the central protostars to grow into SMSs, even in the presence of fragmentation and the formation of numerous low-mass stars.

At [Z/H]$ = -3$, metal-line cooling becomes effective at relatively lower densities ($\sim 10^{6}~\mathrm{cm^{-3}}$), causing the gas temperature to drop earlier during the collapse (Fig. \ref{fig::rhoT_hist}). As a result, filamentary structures on relatively large scales (on the order of parsecs) are observed (Fig. \ref{fig::snap_overall_density}, bottom).
In contrast, focusing on the high-density regions near the center with $\gtrsim 10^{10}~\mathrm{cm^{-3}}$, these regions exhibit a more spherical morphology at an early stage, around $t = 100$ years, extending only a few hundred au (Fig. \ref{fig::snap_evo}, second-right and top panel). Due to this compact, spherical structure, fragmentation does not occur during the initial collapse phase. Instead, a protostar forms at the center, surrounded by a circumstellar disk that begins to develop by $t = 100$ years.
Disk fragmentation subsequently occurs, leading to the formation of multiple stars. By $t = 10^3$ years, this process results in the formation of four to six stars within the disk region.
At this stage, the individual stellar masses remain relatively small, below $10~M_\odot$.
By $t = 10^4$ years, the circumstellar disk evolves into a massive structure, similar to those observed in lower-metallicity cases. This growth occurs as the disk accretes gas from the surrounding envelope, which is continuously replenished by larger-scale, dense ($\gtrsim 10^6~\mathrm{cm^{-3}}$) filamentary streams (bottom panel of Fig.~\ref{fig::snap_overall_density}).
Within this massive disk, a central binary system forms, consisting of two massive stars. The binary system dominates the kinematics of the disk gas through its strong gravitational influence. Consequently, the majority of the gas in the disk is accreted by the central binary stars. Meanwhile, low-mass stars continue to form through fragmentation triggered by metal-line and dust cooling processes within the disk. 
Most of the low-mass stars, however, are either ejected from the disk due to gravitational interactions or merge with the central massive stars. By $t=10^4$ years, the disk has a size of approximately $10^3~$au, and the central protostars have grown to several $100~M_\odot$. This mass is an order of magnitude smaller compared to the central stars in lower-metallicity cases at the same evolutionary stage.
The central stellar mass continues to grow over time, fueled by the dense filament that steadily supplies mass. The accretion rate remains approximately $0.1~M_\odot\mathrm{yr^{-1}}$, and the mass of the central stars ultimately reaches $10^4~M_\odot$ by $t=10^5$ years, as shown in Fig.~\ref{fig::mass_evolution}. This indicates the importance of the accretion from large-scale structures to form SMSs, even at the relatively higher metallicity of [Z/H]$=-3$.

At the highest metallicity considered, [Z/H]$ = -2$, the cloud undergoes significant fragmentation at relatively low densities of $\sim 10^3~\mathrm{cm^{-3}}$ due to efficient metal-line cooling (Fig.~\ref{fig::rhoT_hist}). This results in the formation of multiple dense cores distributed over scales of $\sim 10~$pc (Fig.~\ref{fig::snap_overall_density}, middle panels). Among these, we focus on an isolated core located at the cloud center, with a size of approximately 100 au, where star formation proceeds.
By $t = 100$ years, this central core hosts a single protostar, as shown in Fig.~\ref{fig::snap_evo} (top panel). At this stage, the surrounding circumstellar disk has begun to form, and by $t = 1000$ years, the disk undergoes fragmentation, resulting in the formation of three additional protostars. These stars form a multiple system.
As gas continues to accrete from the cloud envelope, none of the stars in the system grow efficiently, with their masses remaining below $100~M_\odot$ even by $t=10^5$ years. This corresponds to an average accretion rate of approximately $\sim 10^{-3}~M_\odot\mathrm{yr^{-1}}$. 
The low accretion rate in the [Z/H]$=-2$ case is primarily due to the significantly reduced density surrounding the central multiple system, compared to lower-metallicity environments. 
The cloud cores form through fragmentation at a relatively low density of $10^3~\mathrm{cm^{-3}}$, where efficient metal-line cooling and initial turbulence promote fragmentation, which results in a smaller core mass and thus a smaller stellar mass (see Fig.~\ref{fig::snap_overall_density}). 
Another key factor is the absence of substantial gas inflow from larger scales. Unlike in the [Z/H]$=-3$ case, where dense filamentary inflows sustain rapid accretion onto the central stars, fragmentation of the cloud at [Z/H]$=-2$ prevents such large-scale accretion. This prohibits the external gas supply and limits the growth of the central stellar system, resulting in significantly reduced final masses.

To summarize, we have examined how fragmentation at different scales affects the growth of central massive stars, as depicted in Fig.~\ref{fig::mass_evolution}. In the cases with [Z/H]$\leq-4$, where fragmentation occurs only on sub-pc scales, the central stars grow efficiently, ultimately forming SMSs. At [Z/H]$=-3$, pc-scale fragmentation slows the growth of the central stars compared to lower-metallicity cases. Nevertheless, the central stars still reach the SMS regime, albeit with slightly lower masses than those in [Z/H]$\leq-4$. In contrast, at [Z/H]$=-2$, fragmentation on scales of up to 10 pc significantly impedes the growth of the central stars. As a result, their masses barely exceed $1000~M_\odot$, falling short of the SMS range.

Finally, we examine how the most massive stars acquire their mass—whether primarily through gas accretion or via mergers with other protostars, illustrated in Fig.~\ref{fig::mass_by_mergers} as a function of metallicity. 
Panel (a) depicts the ratio of mass gained through stellar mergers by the end of the high-resolution runs, corresponding to $10^4$ years for most cases, except for [Z/H]$=-2$, where it extends to $10^5$ years.
The mass fraction contributed by mergers peaks at [Z/H]$=-4$, where dust cooling is most effective, driving vigorous fragmentation at sub-pc scales. The low-mass stars formed through this fragmentation migrate inward along the accretion flow and eventually merge with the central SMSs, significantly boosting their mass growth.
At metallicities above [Z/H]$\gtrsim -3$, the contribution from mergers decreases. In these environments, the circumstellar disk surrounding the central massive stellar system becomes smaller due to the reduced accretion rate. This smaller disk is more stable and less prone to fragmentation, resulting in fewer low-mass stars available to merge with the central stars. Consequently, the growth of the central massive stars relies more on direct accretion rather than mergers.

Panel (b) presents the same quantities from the long-term, low-resolution runs. Due to the lower resolution, the formation of low-mass stars through small-scale fragmentation and their subsequent mergers with the central massive stars cannot be resolved. 
However, as we will see later, the total mass of these unresolved low-mass stars is relatively small compared to the mass of the central massive stars. Therefore, the lack of resolution does not significantly alter the conclusions. 
The figure shows that, across different metallicities, the mass fraction contributed by mergers remains relatively constant, at around $60$--$70\%$. At [Z/H]$\lesssim -4$, approximately $30$--$40\%$ of the mass originates from mergers with SMSs having masses of $M_* > 10^4~M_\odot$.
This is because, by $\gtrsim 10^5$ years, the available gas is nearly depleted, and the stellar system transitions to a phase dominated by few-body interactions among massive stars, which drive subsequent mergers.
In contrast, the mass fractions contributed by mergers involving very massive stars with $M_* > 10^3$ and $10^4~M_\odot$ decrease for [Z/H]$\gtrsim -3$. This reduction occurs because the central stellar mass becomes smaller as the metallicity increases in this range.
With a smaller central stellar mass, the contribution from mergers with other massive stars diminishes, as the stellar mass must always exceed the mass of the merging companion stars. Consequently, fewer massive companions are available to contribute significantly to the growth of the central star through mergers.

The mass fraction contributed by mergers in our simulations is comparable to or slightly higher than the 40--50\% reported by \citet{Reinoso+2023}, who studied SMS formation in dense gas clouds hosting pre-existing stellar clusters. This slight difference can be attributed to variations in the initial angular momentum of the system.
In our calculations, a large circumbinary disk forms around the SMSs and undergoes fragmentation, producing a hierarchical multiple system. Once the gas supply is depleted, few-body interactions among the resulting stars drive mergers, leading to further mass growth of the SMSs. This process explains the slightly higher merger contribution observed in our results.

\begin{figure}
\centering
\includegraphics[width=0.4\textwidth]{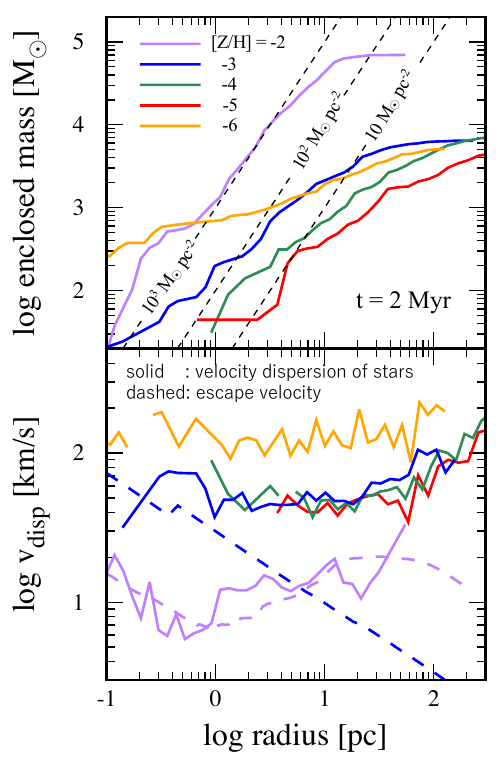}
\caption{
Top panel: Cumulative mass distribution of stars with masses $M_* < 100~M_\odot$ for different metallicities: [Z/H]$=-2$ (purple), [Z/H]$=-3$ (blue), [Z/H]$=-4$ (green), [Z/H]$=-5$ (red), and [Z/H]$=-6$ (yellow) at $t=2~$Myr taken from the long-term, low-resolution runs. Bottom panel: 
Radial profiles of the velocity dispersion for stars with $M_* < 100~M_\odot$ for the same metallicity cases. Dashed lines indicate the escape velocities for the [Z/H]$=-2$ and [Z/H]$=-3$ cases, while those for [Z/H]$\lesssim -4$ are omitted for clarity, as they are comparable to the [Z/H]$=-3$ case. The radii are measured from the center of mass of the VMS/SMSs.
}
\label{fig::star_cluster}
\end{figure}

\subsection{Properties of forming stellar systems} \label{sec::formation_of_stellar_system}

This section presents the properties of the formed stellar systems over various metallicities.
Below, we first overview the structure of star clusters in Section~\ref{sec::results_cluster}, 
present the mass distribution of formed stars
in Section~\ref{sec::mass_spectra} and discuss   
the feedback effects in Section~\ref{sec::early_evolution}
for different metallicities.

\subsubsection{Overall structure of star clusters}
\label{sec::results_cluster}

Fig.~\ref{fig::snap_overall_tMyr} displays the distributions of gas and stars around the clouds at $\sim$Myr after the formation of the first protostar, shortly before supernova feedback will terminate star formation.
The characteristics of the resulting stellar systems vary significantly with metallicity, exhibiting a transition around [Z/H]$=-3$.
At lower metallicities ([Z/H]$\lesssim -3$), SMSs that formed and remain at the cloud center (Sec.~\ref{sec::overall}) are surrounded by a swarm of runaway stars. These stars are dynamically ejected due to interactions with the central SMSs. 
At a higher metallicity of [Z/H]$=-2$, a star cluster with a total stellar mass of approximately $10^5~M_\odot$ forms, spatially extending over several tens of parsecs.

The impact of stellar radiative feedback is clearly evident in the cases with [Z/H]$\lesssim -3$, as the density surrounding the central massive stars significantly decreases.
In these cases, massive stars, sparsely distributed over scales of several tens of parsecs, efficiently photoionize their surroundings. These stars migrated into the less dense outer regions at distances of $\gtrsim 1$--$10~$pc after ejection from the central region due to dynamical interactions with the central SMSs.
While most stars initially form within compact central regions of $\lesssim 0.1~$pc, the majority of them are subsequently ejected through this process. This dynamical ejection has a profound impact on the evolution of the star-forming cloud by accelerating the ionization and evaporation of the surrounding gas. It quenches the mass inflow, halting star formation by approximately 2 Myr.

In contrast, at a higher metallicity of [Z/H]$=-2$, cloud fragmentation occurs on scales exceeding $10~\mathrm{pc}$, resulting in massive stars being distributed across a more extended region. While ionized regions develop around individual massive stars, the cloud retains sufficient mass to gravitationally confine the ionized gas near the central massive star cluster.
This configuration allows star formation to persist until the end of the simulation. Unlike the lower-metallicity cases, where efficient ionization and evaporation truncate star formation, the higher metallicity leads to a reduced number of massive stars and, consequently, a lower overall UV photon emissivity. This diminished radiative output hampers the photoevaporation of the entire cloud, enabling the continued accumulation of gas and prolonged star formation (see Fig.~\ref{fig::resolution} for the time evolution of the total stellar mass).
Note that this behavior agrees with the findings from star cluster formation simulations by \citet{Fukushima+2021}, which demonstrated that the star formation efficiency approaches nearly unity when the initial cloud mass exceeds $10^5~M_\odot$. Star formation will eventually terminate by the impact of supernova explosions.

The structures of star clusters after 2 Myr of evolution, as obtained from the long-term, low-resolution runs, are shown in Fig.~\ref{fig::star_cluster}. The figure illustrates (a) the total stellar mass enclosed within a given radius and (b) the velocity dispersion of stars as functions of the distance from the center for different metallicities. To ensure the analysis focuses on the general properties of the clusters, the contribution of SMSs and VMSs is excluded by limiting the sample to stars with masses below $100~M_\odot$.
At metallicities below [Z/H]$\simeq -3$, the star clusters exhibit similar total masses and velocity distributions, with the total stellar mass reaching up to $10^4~M_\odot$ and spatially extending to roughly 100 pc. The velocity dispersion exceeds several tens of $\mathrm{km}~\mathrm{s^{-1}}$, indicating that many stars are escaping from the cluster due to dynamical interactions with the central massive stars. Indeed, most stars initially form within circumstellar disks located within 0.1 pc of the central massive stars, where they are particularly susceptible to ejection caused by gravitational interactions. 
In contrast, at [Z/H]$=-2$, the radial structure of the cluster differs significantly from the lower-metallicity cases. A massive stellar system, with a total mass of $10^4$--$10^5~M_\odot$, is concentrated within the central several parsecs and exhibits a lower velocity dispersion of approximately $10~\mathrm{kms^{-1}}$, indicating that the system is gravitationally bound. The size of this cluster is determined by the Jeans length at the point where metal-line cooling becomes effective, leading to a significant drop in temperature (Fig.~\ref{fig::rhoT_hist}). Compared to cases without external FUV radiation, the presence of the FUV field delays the temperature drop to higher densities. As a result, the resulting star cluster becomes more compact and denser due to this delayed cooling effect.
Such systems are likely to evolve into dense star clusters, such as globular clusters observed in the local universe, after the eventual death of the central SMSs. This point will be discussed further in Section~\ref{sec::discussion}.

\begin{figure*}
\centering
\includegraphics[width=0.95\textwidth]{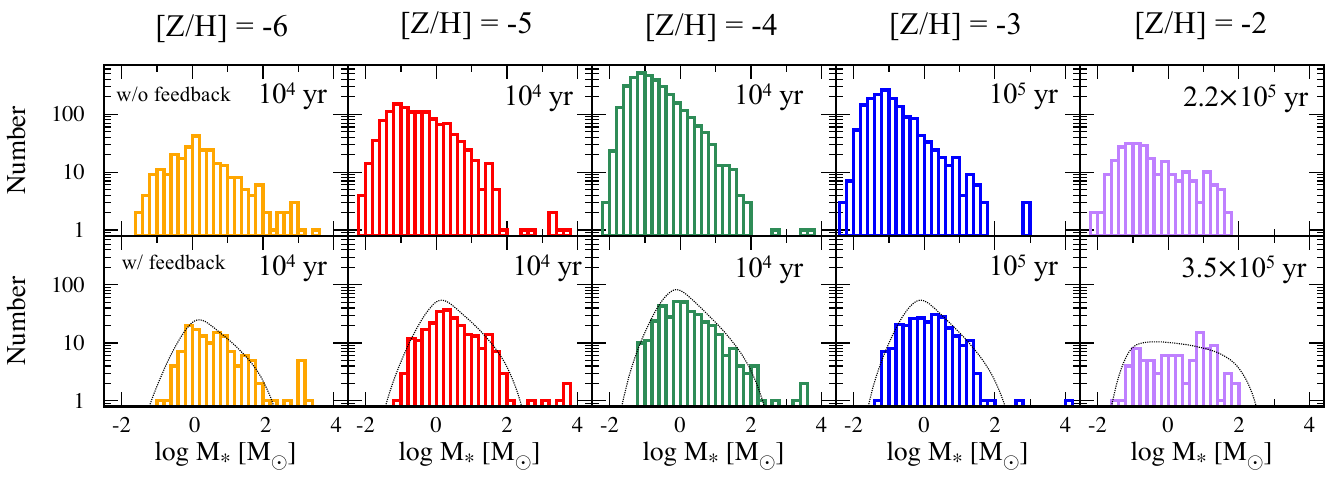}
\caption{
The mass distributions at the end of the high-resolution simulation for [Z/H]$=-6$, $-5$, $-4$, $-3$, and $-2$ from left to right columns.
The top and bottom panels show the distribution when we do not include and include stellar feedback.
We also attach the time at the end of high-resolution runs after the formation of the first protostar.
The dotted lines in the bottom panels show the fitted spectrum (equation~\ref{eq::IMF_tapered}) for each distribution whose parameters are listed in table~\ref{tab::IMF_fit}.
}
\label{fig::mass_spectra}
\end{figure*}

\begin{figure*}
\centering
\includegraphics[width=0.95\textwidth]{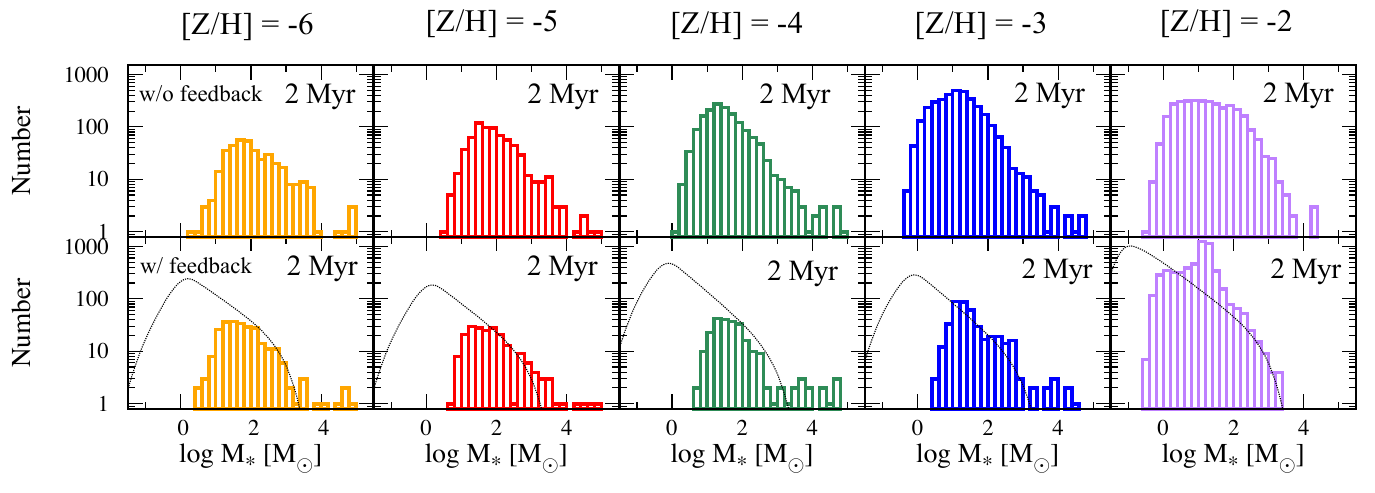}
\caption{
The same as Fig.~\ref{fig::mass_spectra} but for low-resolution runs. The dotted lines in the bottom panels show the extrapolation of the mass distribution expected from the high-resolution runs, with a form of equation~\ref{eq::IMF_tapered}.
}
\label{fig::mass_spectra_t2Myr}
\end{figure*}

\subsubsection{Mass distributions of individual stars} \label{sec::mass_spectra}
Here, we present the resulting stellar mass spectra for different metallicities. To construct a complete mass spectrum, we combine the results from both the short-term, high-resolution runs and the long-term, low-resolution runs.
The short-term, high-resolution runs, which track the evolution during the relatively early stages, spanning from $10^4$ to $10^5$ years, resolve the protostellar radii of accreting SMSs and allow for the study of very low-mass objects, down to $\sim 0.01~M_{\odot}$. However, these runs fail to capture the full epoch of star formation, particularly the late-stage evolution, leading to an underestimation of the final masses of SMSs.
To address this limitation, we also perform long-term, low-resolution runs, which extend up to 2 Myr—the timescale when supernovae are expected to disrupt the cloud and halt star formation. While the lower resolution limits our ability to examine the distribution of very low-mass objects, it effectively captures the growth of massive stars, including the final masses of SMSs.
By reconstructing the missing low-mass component based on the short-term, high-resolution results, we construct a comprehensive distribution of stellar masses, integrating the contributions from both the early and late stages of star formation.

We first present the stellar mass spectra at the end of our short-term, high-resolution simulations in Fig.~\ref{fig::mass_spectra}. A comparison between the cases without stellar feedback (top panel) and those with stellar feedback (bottom panel) reveals that feedback mechanisms primarily influence the distribution of low-mass stars. In contrast, the distribution of massive stars remains largely unaffected by the inclusion of feedback.

Without stellar feedback (top panels of Fig.~\ref{fig::mass_spectra}), the numbers and distributions of low-mass stars exhibit a strong dependence on metallicity.
At the lowest metallicity, [Z/H]$=-6$, where metallicity effects are nearly negligible, the mass spectrum is characterized by a roughly log-flat distribution with a minor peak around $1~M_\odot$. As metallicity increases to [Z/H]$=-5$ and $-4$, low-mass stars predominantly form through fragmentation triggered by dust cooling. Consequently, the number of low-mass stars increases, while the peak of the mass spectrum consistently remains at approximately $0.1~M_\odot$.
For higher metallicities, [Z/H]$=-3$ and $-2$, the number of low-mass stars decreases compared to the [Z/H]$=-4$ case. This reduction is attributed to a lower total gas mass available for star formation and, consequently, a decrease in the total stellar mass, driven by the reduced accretion rate (see Fig.~\ref{fig::mdot}).

When radiative feedback is included, the number of low-mass stars is significantly reduced across all metallicities. The resulting mass spectra for very low metallicities, [Z/H]$\lesssim -3$, exhibit a similar shape, with a peak around $1~M_\odot$. The reduction in the number of stars with $M_* \lesssim 0.1~M_\odot$ is attributed to the suppression of small-scale fragmentation by radiative heating of dust grains, further discussed in Section~\ref{sec::early_evolution}.
For [Z/H]$\lesssim -3$, the mass spectra can be decomposed into two parts: a low-mass component resembling a Chabrier-like distribution, peaking at approximately $1~M_\odot$ and extending up to around $100~M_\odot$ \citep{Chabrier2003}, and an SMS component with masses ranging from $10^3$ to $10^4~M_\odot$. The latter dominates the total stellar mass and consists of a hierarchical multiple system.
While the fraction of mass contributed by the SMS component remains comparable between cases with and without feedback across all metallicities, variations in the detailed mass distribution primarily result from stochastic few-body interactions.
In contrast, the mass spectrum for [Z/H]$=-2$ is markedly different, consisting solely of a low-mass population with a log-flat distribution. This stark contrast shows a sharp transition in the mass spectra between metallicities of [Z/H]$=-2$ and $-3$.

\begin{table}
    \centering
    \begin{tabular}{c||c|c|c|c|c}
        [Z/H] & -6 & -5 & -4 & -3 & -2  \\
        \hline
        $\alpha$ & 1.46 & 1.51 & 1.57 & 1.57 & 1.09 \\
        $m_0 / M_\odot$ & 0.98 & 0.93 & 0.53 & 0.53 & 0.05  
    \end{tabular}
    \caption{The best fit IMF parameters in equation~\ref{eq::IMF_tapered} for different metallicities. 
    $\alpha$: the slope at the high-mass end, $m_0$: the peak mass.}
    \label{tab::IMF_fit}
\end{table}

To quantitatively compare the mass spectra across different metallicities, we fit the low-mass Chabrier-like component using the "tapered power-law" model described by \citet{deMarchi+2005}:
\begin{align} \label{eq::IMF_tapered}
\phi(M_*) = \phi_0 M_*^{-\alpha} \left[ 1 - \exp \left(-\left( \frac{M_*}{m_0}\right)^{1.6}\right)\right] 
\exp\left(-\frac{m_\text{min}}{M_*}-\frac{M_*}{m_\text{max}}\right),
\end{align}
where the parameters $M_\text{max}=150~M_\odot$ and $M_\text{min}=0.04~M_\odot$ are fixed constants.
Three free parameters are adjusted to fit the data: $\phi_0$, a normalization constant, $m_0$, the peak mass of the distribution, and $\alpha$, the slope of the high-mass end. The best-fit parameters for each metallicity are summarized in Table~\ref{tab::IMF_fit}. 
This approach provides a robust comparison of the low-mass stellar population across the metallicity range and enables the reconstruction of low-mass components unresolved in the long-term, low-resolution runs using the short-term, high-resolution results.

At metallicities of [Z/H]$\lesssim -3$, the fitted slope $\alpha$ is approximately 1.5, showing little dependence on metallicity. This value is notably shallower than the Salpeter slope of $\alpha=2.35$, indicating a relatively top-heavy distribution even within the low-mass stellar population. 
Additionally, the peak mass, 
$m_0$, decreases with increasing metallicity, as enhanced dust cooling at higher metallicities promotes fragmentation by lowering the Jeans mass, thereby reducing the typical mass of the resulting fragments.
In contrast, at [Z/H]$=-2$, the mass spectrum is best described by a single power-law distribution with $\alpha=1.1$, representing a nearly log-flat IMF. 

The mass spectra of stellar clusters 2 Myr after the formation of the first protostar, derived from the long-term, low-resolution runs, are presented in Fig.~\ref{fig::mass_spectra_t2Myr}. As before, the results for cases without and with stellar feedback are shown in the top and bottom panels, respectively.

In the absence of feedback, the spectra extend to the high-mass range of $10^4$--$10^5~M_\odot$ across all metallicities. The number of non-VMS/SMS stars with $M_* \lesssim 1000~M_\odot$, peaking at approximately $10~M_\odot$, increases with metallicity. This trend can be attributed to more vigorous fragmentation facilitated by the enhanced cooling efficiency at higher metallicities.

When feedback is included, a significant reduction is observed in the population of low-mass stars ($M_* \lesssim 10~M_\odot$) for metallicities of [Z/H]$\lesssim -3$, primarily due to heating of dust grains and suppression of small-scale fragmentation. However, the high-mass end of the spectra, particularly for $M_* \gtrsim 10^4~M_\odot$, remains unaffected. This indicates that feedback has a minimal impact on the growth of SMSs.

For metallicities below [Z/H]$\simeq -3$, the resulting spectra are similar, comprising a low-mass population peaking around $10~M_\odot$ and an SMS component extending to $10^4$--$10^5~M_\odot$. It is worth noting that the distribution near the low-mass peak is sensitive to numerical resolution; the higher-resolution simulations produce spectra that extend further into lower stellar masses. This suggests that the actual mass spectra might extend even further into the low-mass range.

Unlike the lower-metallicity cases, at [Z/H]$=-2$, stellar feedback significantly impacts the high-mass end of the spectrum while having minimal influence on the lower-mass end. The growth of stars with $M_* \gtrsim 10~M_\odot$ is strongly suppressed by ionizing radiation from the forming stars. This is similar to the present-day star formation, where ionizing radiation from accreting massive protostars disrupts accretion flows, effectively halting their growth \citep[e.g.,][]{Bate2009}.
At this metallicity, typical accretion rates in our calculation fall below the critical threshold of $0.04~M_\odot \mathrm{yr^{-1}}$, required to maintain stars in an inflated supergiant protostar state \citep{Hosokawa+2011}. As a result, the stars contract to the main-sequence phase, where ionizing radiation feedback operates more efficiently, quenching further mass growth.
The shape of the mass spectrum at the lower-mass end remains largely unchanged because the star-forming region extends over several tens of parsecs, whereas dust heating, which can suppress the formation of stars below $1~M_\odot$, is limited to more compact regions. However, this effect is less evident in the low-resolution simulations, as it falls below the resolution limit, rendering its impact on the mass spectrum negligible in this context.

To address the missing low-mass stars in the low-resolution runs due to limited resolution, we reconstruct the stellar population at the lower-mass end using predictions from the high-resolution runs. Specifically, we extrapolate the stellar mass distribution for $M_* \lesssim 10M_\odot$ by employing the functional form of the fitted IMFs derived from the high-resolution runs (as described in Equation\ref{eq::IMF_tapered}), 
except for the case of [Z/H]$=-2$.
For normalization, we use the number of higher-mass stars within the range $10 < M_*/M_\odot < 1000$.

For [Z/H]$=-2$, we adopt a slope $\alpha$ of 1.5 instead of the value of 1 obtained from the fitting, as this steeper slope aligns more closely with trends often observed in lower-metallicity cases. Extrapolation from the high-resolution run is not feasible in this instance due to the differing origins of low-mass stars in the two types of simulations. In the high-resolution run, low-mass stars predominantly form through fragmentation within a few star-forming disks. However, in the low-resolution run, low-mass stars emerge from fragmentation over more spatially extended regions, driven by turbulent flows combined with efficient line cooling. Consequently, the slope of the mass spectrum in the low-resolution run is expected to be significantly steeper than $\alpha=1$, which is typically observed in high-resolution runs \citep[e.g.][]{Hennebelle+2008}.

The reconstructed population is represented by the dotted lines in Fig.~\ref{fig::mass_spectra_t2Myr}, which aligns well with the distribution observed at the high-mass end. This approach ensures a more comprehensive representation of the stellar mass spectrum, bridging the gap caused by resolution limitations in the low-resolution simulations.

In all cases, the missing low-mass population dominates in terms of the number of stars but contributes less than $1\%$ to the total stellar mass. Therefore, the presence of these low-mass stars has negligible impact on the star cluster formation process and its subsequent evolution. As a result, the overall reliability of the low-resolution runs, particularly for the higher-mass end of the mass spectrum, remains reliable and robust.

\begin{figure*}
\centering
\includegraphics[width=0.95\textwidth]{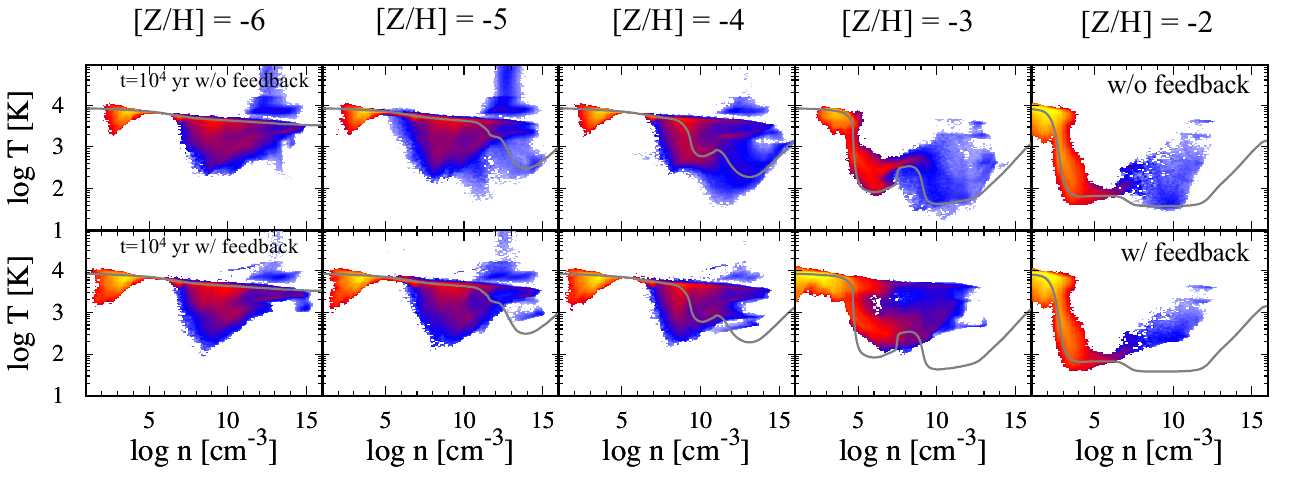}
\caption{
Distributions of gas density and temperature at $10^4~$yr after the formation of the first protostar, same as in Fig.~\ref{fig::rhoT_hist}. The top panels represent cases with stellar feedback included, while the bottom panels show cases without stellar feedback. This comparison illustrates the influence of stellar feedback on the thermal and density evolution of the gas during the early stages of star formation.
}
\label{fig::rhoT_hist_t1e4}
\end{figure*}

\subsubsection{Feedback effects on the final stellar systems} \label{sec::early_evolution}

\begin{figure}
\centering
\includegraphics[width=0.5\textwidth]{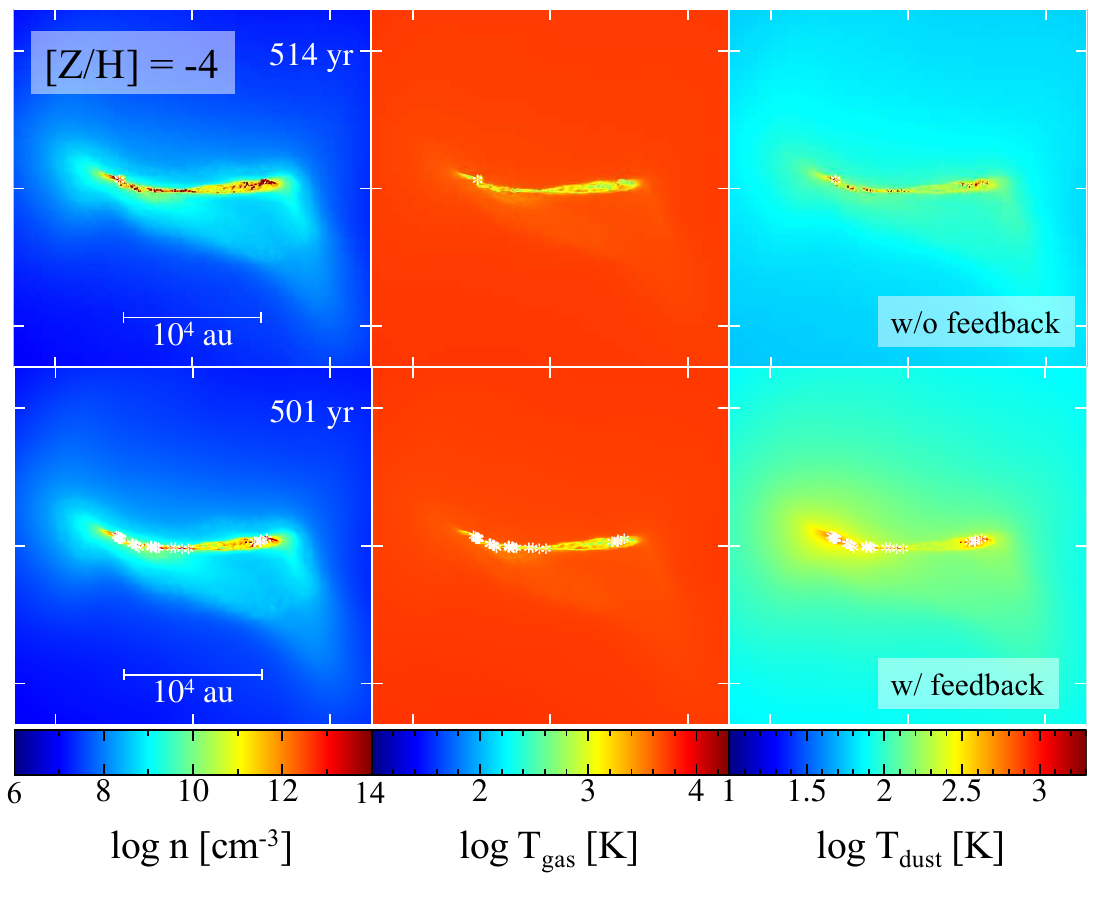}
\caption{
Projected distributions of density (left), gas temperature (middle), and dust temperature (right) for the case of [Z/H]$=-4$. The top panels represent the case with stellar feedback included, while the bottom panels show that without stellar feedback. Asterisks mark the locations of protostars. 
}
\label{fig::snap_M-4}
\end{figure}

\begin{figure}
\centering
\includegraphics[width=0.5\textwidth]{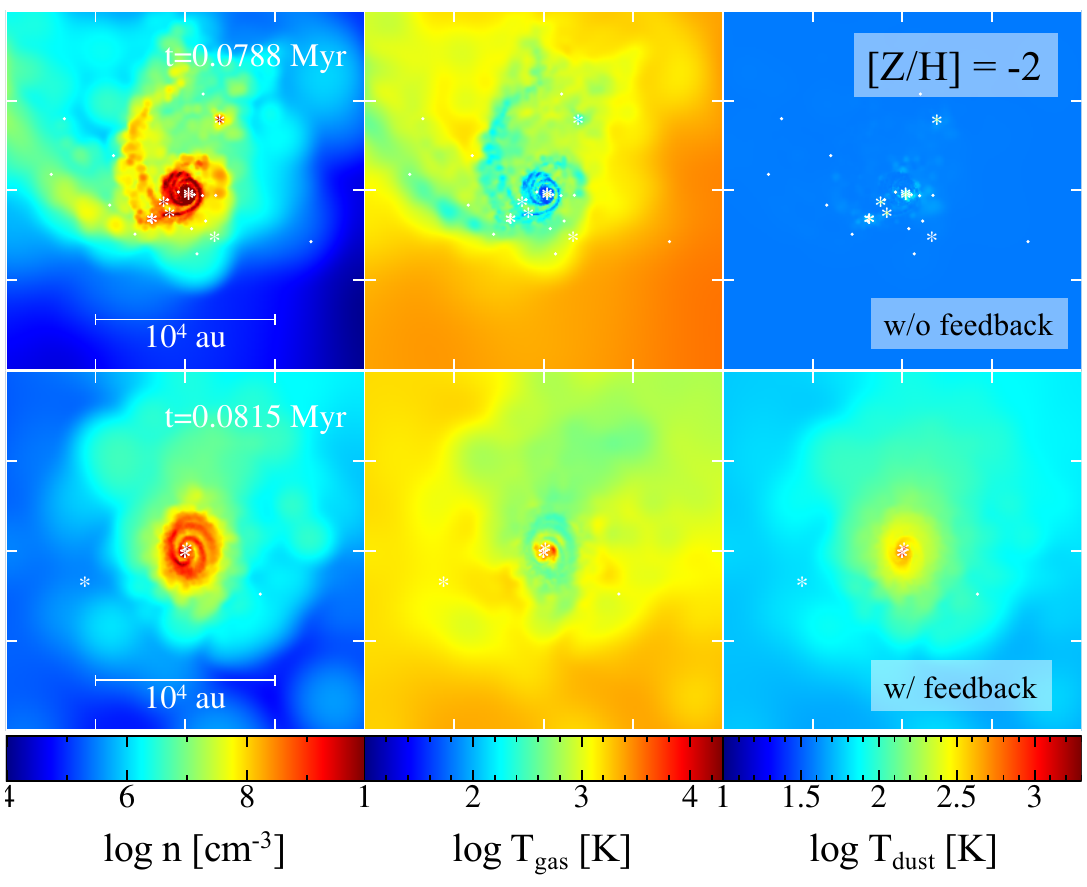}
\caption{
Projected distributions of density (left), gas temperature (middle), and dust temperature (right) for the case of [Z/H]$=-2$, similar to Fig.~\ref{fig::snap_M-4}. 
}
\label{fig::snap_M-2}
\end{figure}

To understand the origin of the typical stellar masses of $\sim 1~M_\odot$ (Fig.~\ref{fig::mass_spectra}), we analyze the thermal evolution of the gas during the accretion phase of growing stars in clusters.
Fig.~\ref{fig::rhoT_hist_t1e4} presents the gas distributions in the density-temperature plot at $10^4$ years after the formation of the first protostar, comparing cases with (top panels) and without (bottom panels) feedback effects.

Once protostars form, the temperature distribution deviates significantly from that observed during the initial collapse phase (Fig.~\ref{fig::rhoT_hist}), reflecting the influence of radiative feedback due to ongoing star formation processes.

With slight metallicity, i.e., for [Z/H]$\gtrsim -5$, the temperature drops to several hundred Kelvin due to dust cooling in cases without radiative feedback. When radiative feedback is included, it heats the dust, suppressing its cooling effect and resulting in similar temperature distributions across cases with metallicities below [Z/H]$\sim -4$. At densities of approximately $10^7~\mathrm{cm^{-3}}$, the temperature in these cases suddenly decreases to around $100~$K, although such cold gas component
contributes minimally to the total mass.
Interestingly, this temperature drop is observed even in the case of [Z/H]$=-6$, where dust cooling is negligible. This phenomenon arises from a two-stage process \citep[e.g.,][]{Inayoshi+2014, Matsukoba+2021, Prole+2024}:
gas behind a spiral arm, which forms due to gravitational instability in a circum-stellar/binary disk (Fig.~\ref{fig::snap_evo}), 
undergoes expansion and cools adiabatically.
In this cooled gas, H$_2$ formation is enhanced, further increasing the cooling efficiency and generating the cold component.
This cold component is prone to fragmentation, leading to the formation of low-mass stars, even at the lowest metallicity of [Z/H]$=-6$, as shown in Fig.~\ref{fig::snap_evo}.

For [Z/H]$\gtrsim -3$, radiative feedback has minimal impact on the temperature evolution in the low-density regime ($n\lesssim 10^9~\mathrm{cm^{-3}}$), where metal-line cooling is the primary cooling mechanism. However, at higher densities, where dust cooling becomes the dominant process, radiative feedback substantially suppresses cooling and increases the gas temperature.

Stellar irradiation affects the dynamics of circumstellar disks by raising the dust temperature, thereby suppressing fragmentation \citep{Bate2009, Chon+2024}. To investigate how dust heating reduces small-scale fragmentation and modifies the typical fragmentation mass scale, we analyze its effects in detail. Fig.~\ref{fig::snap_M-4} illustrates the role of dust cooling in driving fragmentation in the absence of feedback, and the subsequent modification of this process by radiative feedback for [Z/H]$=-4$. This case serves as a representative example for the very low-metallicity range of $-5 \lesssim$ [Z/H] $\lesssim -3$, where dust cooling causes the temperature to drop at densities above $10^{10}~\mathrm{cm^{-3}}$.

In both cases, with and without feedback, filamentary structures form and fragment into low-mass stars. This indicates that dust heating does not entirely prevent fragmentation on scales of $100$--$1000$ au within filaments during the early phase of evolution ($\lesssim 10^3~$years).
In the later phase (around $\sim 10^4$ years), as the central star grows to a mass of $10^4M_\odot$, dust heating suppresses fragmentation on scales of $100$--$1000$ au throughout the entire region. During this stage, a rotationally-supported disk develops around the central massive stars (Fig.~\ref{fig::snap_evo}).
Irradiation from these massive stars raises the temperature of the disk gas, effectively disabling dust cooling within the disk and further stabilizing it against fragmentation.

This stabilization effect is evident in the density-temperature plot shown in Fig.~\ref{fig::rhoT_hist_t1e4}. Stellar irradiation raises the dust temperature, which in turn increases the gas temperature. For [Z/H]$\lesssim -4$, where dust grains serve as the primary coolant at densities around $10^{10}~\mathrm{cm^{-3}}$, the temperature of the cold gas component, typically dropping below 100 K due to dust cooling in the absence of feedback, rises to several hundred K under the influence of irradiation.
A similar effect is observed for [Z/H]$=-3$ and $-5$, where stellar radiation globally heats the gas, effectively suppressing dust cooling. In these cases, the gas is no longer able to cool efficiently, preventing the formation of dense, cold regions conducive to fragmentation.

As a result, the mass spectra for these metallicities resemble that of [Z/H]$=-6$, where dust cooling is negligible, and fragmentation is inherently limited (Fig.~\ref{fig::mass_spectra}). This highlights the significant role of radiative feedback in homogenizing the star-forming environments across varying metallicities by suppressing small-scale fragmentation.

As the metallicity increases, the influence of dust heating becomes more pronounced, effectively suppressing fragmentation within the stellar surroundings. Fig.~\ref{fig::snap_M-2} illustrates how radiative feedback stabilizes the circumstellar disk, newly formed around the most massive star, for [Z/H]$=-2$ (see also Fig.~\ref{fig::snap_evo}).

Without radiative feedback, the gas temperature in the disk at densities of $\gtrsim 10^6~\mathrm{cm^{-3}}$ drops to several tens of Kelvin due to efficient cooling. In contrast, with feedback, the temperature remains above 100 K (see also Fig.~\ref{fig::rhoT_hist_t1e4}). The heating effect is particularly significant in this case because the disk is smaller, and the dust temperature is generally lower than in lower-metallicity cases.
This elevated gas temperature increases the Jeans mass, resulting in larger-scale fragmentation and higher typical stellar masses. Consequently, radiative feedback not only stabilizes the disk but also alters the star formation process by shifting the mass spectrum toward more massive stars.

In summary, the typical stellar mass of $\sim 1~M_\odot$ arises from the stellar irradiation. The fragmentation scale and typical stellar mass are influenced by metallicity primarily during the early stages of star formation. However, over the entire $\sim$ Myr duration of star formation, the fragmentation mass scale is predominantly governed by stellar irradiation rather than metallicity. As a result, the final mass spectra of stars within clusters exhibit rather weak dependence on metallicity, with a consistent peak mass of approximately $1~M_\odot$ across the various cases.

\section{Summary and Discussion} \label{sec::discussion}
In this study, we investigated the formation and evolution of supermassive stars (SMSs) and star clusters in metal-poor clouds under strong far-ultraviolet (FUV) radiation fields. Using three-dimensional hydrodynamic simulations, we accounted for a detailed chemical network, non-equilibrium thermal evolution, and the effects of radiative feedback from massive stars. Our calculations span a wide range of metallicities, from [Z/H]$=-6$ to $-2$, allowing us to explore the transition from SMS-dominated systems to compact star clusters. By performing both short-term, high-resolution runs and long-term, low-resolution runs, we captured the early stages of fragmentation and accretion, as well as the late-stage evolution of the stellar systems over 2 Myr.

We found that metallicity significantly influences the fragmentation properties and the growth of massive stars. At very low metallicities ([Z/H]$\lesssim -3$), SMSs with $10^4$--$10^5~M_\odot$ can form, albeit through different pathways. In the [Z/H]$=-6$ case, where dust cooling is negligible, the gas collapses nearly monolithically, giving rise to SMSs at the center. For slightly higher metallicities ([Z/H]$=-5$ and $-4$), dust cooling induces sub-pc-scale fragmentation, producing multiple fragments around the central region. However, these fragments are ultimately accreted onto the central protostar or merge with it, allowing the central massive objects to grow into SMSs. Despite initial fragmentation, the mass converges toward the central massive stars, sustaining their efficient growth. For [Z/H]$=-3$, while pc-scale fragmentation slows the mass growth of central stars during the early phases, continuous accretion from dense filaments allows the formation of the SMSs of $\sim 10^4~M_\odot$. At [Z/H]$=-2$, however, large-scale fragmentation on $\sim 10~$pc scales leads to a more spatially-extended star formation, resulting in the formation of a massive star cluster with a central very massive star (VMS) of a few $10^3~M_\odot$.

Radiative feedback plays a critical role in regulating fragmentation and the growth of massive stars. At metallicities of [Z/H]$\lesssim -3$, ionizing radiation is largely ineffective due to the high-density environments surrounding the central stars, allowing the formation of SMSs. Dust heating by stellar irradiation becomes significant at later stages, around $10^4~$yr after the onset of star formation, effectively suppressing the formation of low-mass stars within the circumstellar disk. As a result, the stellar mass distribution at [Z/H]$\lesssim -3$ becomes nearly indistinguishable across different metallicities.
For higher metallicities of [Z/H]$=-2$, the effect of radiative feedback, particularly ionizing radiation, is more pronounced. It significantly alters the dynamics of the surrounding gas, quenching the mass accretion onto the central stars. The combination of lower accretion rates, caused by large-scale fragmentation, and the enhanced efficiency of ionizing radiation, further suppresses the growth of massive stars. Consequently, the central stellar mass at [Z/H]$=-2$ reaches only $2000~M_\odot$, substantially lower than the SMS masses achieved at lower metallicities.

The resultant mass spectra of star clusters reflect the underlying fragmentation dynamics and feedback effects. For metallicities of [Z/H]$\lesssim -3$, the spectra exhibit a composite distribution: a low-mass Chabrier-like component formed through small-scale fragmentation and an SMS component with masses of $10^4$--$10^5~M_\odot$, which occupies the high-mass end. These clusters are characterized by SMSs at the center, surrounded by low-mass stars that are dynamically ejected from the cluster due to interactions with the central massive objects. At [Z/H]$=-2$, instead of an SMS-dominated structure, a bound star cluster forms with a central VMS of approximately $1000~M_\odot$. 

These findings provide a comprehensive view of the interplay between metallicity, radiative feedback, and accretion dynamics in shaping the stellar populations in metal-poor environments. Our results highlight the importance of metallicity to distinguish the formation of SMSs and dense star clusters, offering new insights to form SMBHs and compact star clusters in the early universe.

~

In our simulations, the masses of the formed stars and their remnant BHs range from $3$ to $8\times 10^4~M_\odot$ for metallicities of [Z/H]$\lesssim -3$. The final mass is determined when the efficient gas supply, at a rate of $\sim 1~M_\odot~\mathrm{yr}^{-1}$, ceases after approximately $10^5~$ years. This quenching of the accretion flow is attributed to the tidal fields of nearby massive galaxies, which disrupt the outer envelope of the cloud and limit the maximum mass that can be supplied to the central stellar system \citep{Chon+2016}. Such a scenario is common in the conventional direct collapse scenario, where SMS-forming clouds are typically located in the vicinity of massive, luminous galaxies and are subjected to strong tidal forces.
On the other hand, when SMS formation occurs in somewhat metal-enriched environments, the cloud may not always experience such strong tidal fields. For instance, a cloud located in a pocket of low-metallicity regions within a halo could receive FUV radiation from diffusely distributed massive stars within the halo. In this case, the reduced tidal forces could enable the formation of even more massive seed BHs.
\citet{Regan+2020} identified halos with high infall rates ($\gtrsim 0.1~M_\odot \mathrm{yr}^{-1}$) as promising sites for SMS formation from a halo sample of the Renaissance simulation. Interestingly, they found that halos capable of sustaining high accretion rates ($\sim 1~M_\odot~\mathrm{yr}^{-1}$) and forming large seed BHs are predominantly those with finite metallicities in the range [Z/H]$\lesssim -5$ to $-3$, rather than pristine, primordial environments. This suggests that the super-competitive accretion scenario in metal-enriched environments offers more opportunities for forming massive seed BHs compared to the conventional direct collapse scenario.

\begin{figure}
\centering
\includegraphics[width=0.47\textwidth]{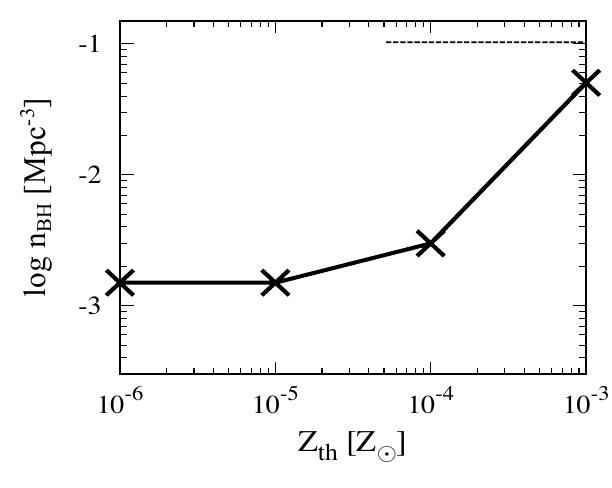}
\caption{
The number density of massive seed BHs as a function of the threshold metallicity below which SMSs can form. The BH densities are derived from the dark matter-only semi-analytic calculations by \citet{Chiaki+2023}, corrected by assuming that only 5\% of the candidate halos identified by \citet{Chiaki+2023} successfully collapse to form massive seed BHs. This success rate is based on \citet{Chon+2016}, where it was shown that cloud collapse within candidate halos is often inhibited by tidal forces from nearby galaxies. The crosses represent the resulting number densities after applying the correction factor, and the horizontal dashed line indicates the number density of SMBHs in the local universe.
}
\label{fig::DCBH_density}
\end{figure}

\begin{figure*}
\centering
\includegraphics[width=1.0\textwidth]{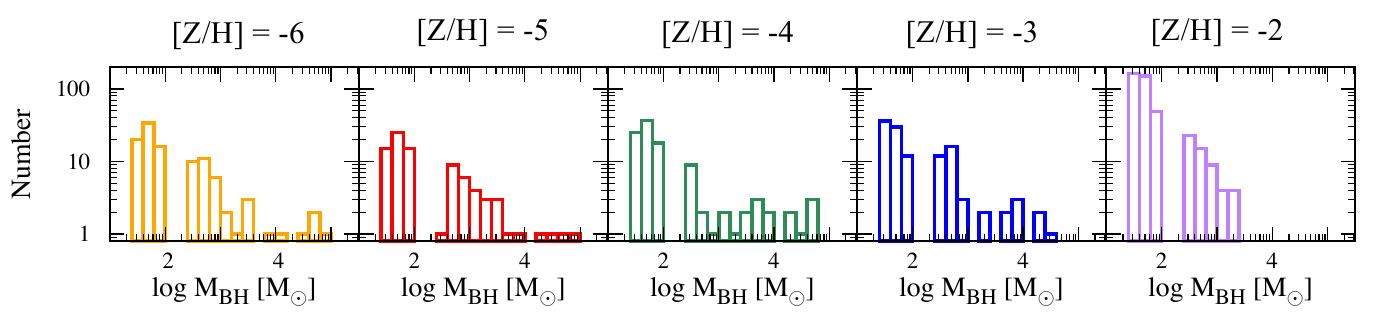}
\caption{
The mass function of remnant BHs in different metallicity environments. Stars with masses in the range $30M_\odot < M_* < 80M_\odot$ or $M_* > 260~M_\odot$ are assumed to directly collapse into BHs, following \citet{HegerWoosley2002}. No mass loss processes are considered in this calculation.
}
\label{fig::BH_mass_function}
\end{figure*}

The ability to form massive seed BHs in environments with [Z/H]$\lesssim -3$ suggests a higher seed number density compared to predictions from the conventional direct collapse scenario, which assumes SMS formation in primordial, pristine gas. A single supernova event typically enriches the surrounding gas to [Z/H]$\sim -6$--$-3$ \citep{Chiaki+2018}, implying that massive seed BHs can form in halos that have undergone a previous episode of star formation.
Cosmological simulations of galaxy formation have shown that during the early stages of galaxy assembly at $z \gtrsim 10$, the typical metallicity of the interstellar medium remains in the range [Z/H]$\sim -4$--$-2$ \citep{Wise+2012, Ricotti+2016}. Recent work by \citet{Sugimura+2024} further suggests that the typical metallicity in early galaxies may be even lower than previously estimated when the suppression of star formation by internal FUV radiation is accounted for. This radiation slows down early star formation and allows for the continued accretion of primordial gas into the halo, keeping the typical gas metallicity around [Z/H]$\sim -3$ for the first $100~$Myr.
\citet{Venditti+2023} have found that even in the galaxy with $M_* \gtrsim 10^8~M_\odot$, a pocket of a metal free or low-metallicity region exist to form stars.

The super-competitive accretion scenario, which allows for SMS formation in gas with finite metallicity, raises an intriguing question: by how much does the number density of massive seed BHs increase compared to the conventional direct collapse scenario? \citet{Chiaki+2023} addressed this question by estimating the number density of seed BHs in environments permitting massive seed BH formation at finite metallicities. They conducted semi-analytic galaxy formation simulations that track the evolution of luminous galaxies and the enrichment of metals within halos.
The resulting number density of massive seed BHs, as a function of the threshold metallicity below which these BHs can form, is shown in Fig.~\ref{fig::DCBH_density}. It is important to note that the number density derived in \citet{Chiaki+2023} represents the density of candidate halos potentially capable of forming seed BHs.
In reality, the seed BH number density is expected to be smaller because the collapse of clouds within these halos is often hindered by tidal effects from nearby massive halos \citep{Chon+2016}. For instance, hydrodynamical simulations by \citet{Chon+2016} found that only 5\% of candidate halos successfully underwent collapse to form massive BHs, with just two out of 42 candidate halos achieving successful collapse.
To account for this, a correction factor of 5\% has been applied to the values reported by \citet{Chiaki+2023}.
The figure indicates that the density of massive seed BHs increases significantly, reaching $0.1$--$1~\mathrm{Mpc^{-3}}$ if SMSs can form in halos with metallicities as high as [Z/H]$\gtrsim -3$. This number density is comparable to that of SMBHs observed in the local universe.
These findings suggest that the super-competitive accretion scenario has the potential to account for the origin of all SMBHs in the universe, not only the highest-redshift ones, from the perspective of their number density, effectively bridging the gap between early seed formation and the observed SMBH population in the local universe.
Our results also provide a theoretical foundation for seeding BHs in low-mass, metal-poor halos, as commonly assumed in cosmological simulations \citep[e.g.][]{Weinberger+2018,Trebitsch+2021,Bhowmick+2024}.

As explored in this study, FUV-irradiated low-metallicity clouds not only produce central massive seed BHs with masses of $10^4$--$10^5~M_\odot$, but also generate a population of intermediate-mass black holes (IMBHs) with masses around $\sim 1000~M_\odot$. Fig.~\ref{fig::BH_mass_function} presents the mass function of the remnant BHs predicted from our simulations. For this analysis, we assume that stars with masses in the range $30~M_\odot < M_* < 80~M_\odot$ or $M_* > 260~M_\odot$ directly collapse into BHs, avoiding pair-instability supernovae \citep{HegerWoosley2002}.
For metallicities of [Z/H]$\lesssim -3$, our simulations show the formation of four to six massive seed BHs, characterized by a log-flat mass spectrum. Additionally, a distinct population of IMBHs emerges with masses below a few thousand $M_\odot$, following a power-law distribution.
In the central region of the cloud, massive BHs typically form binaries or hierarchical multiple systems, while the majority of IMBHs are dynamically ejected due to interactions with the central massive objects. The separations of the central BH binaries are typically $\gtrsim 10^3~$au, resulting in gravitational wave (GW) merger timescales that far exceed the current Hubble time. However, interactions with surrounding gas and stars may assist in shrinking the binary orbits, potentially leading to GW-driven mergers on observable timescales \citep{Mukherjee+2024}.
Such merger events, if they occur, could be detectable by future missions like LiSA, even at high redshifts of up to a few tens. If the formation of massive seed BHs is consistently accompanied by binary formation, future LiSA observations could provide critical constraints on the number density and formation environments of seed BHs \citep{Chon+2018, Hartwig+2018}. These insights would be instrumental in understanding the role of seed BHs in early cosmic structure formation and their contribution to the growth of supermassive black holes in the present-day universe.

Several effects not considered in our current calculations could influence SMS formation. These include the impact of magnetic fields and the effects of mass loss during stellar mergers or due to stellar winds, both of which remain uncertain.

Magnetic fields, absent in our current analyses, may play a critical role in enhancing the masses of forming stars and facilitating SMS formation. These fields can be generated even in chemically pristine environments, such as at accretion shocks in halos, and can be amplified during the gravitational collapse of clouds \citep{McKee2020, Stacy+2022, Sadanari+2023, Higashi+2024, Diaz+2024}. Additionally, stellar feedback, including interactions between seed magnetic fields and expanding supernova shells, can further amplify magnetic fields, especially in metal-enriched environments \citep{Koh+2016}.
Magnetic fields extract angular momentum from accreting gas and suppress fragmentation within circumstellar disks \citep{Hirano+2023, Latif+2023}. This suppression reduces the multiplicity of the central massive stellar system, concentrating the mass into a single central object. As a result, magnetic fields may enable the growth of stars to the SMS regime, reaching masses of $\gtrsim 10^5~M_\odot$.

Another factor not accounted for in our simulations is mass loss, which can occur during mergers or through stellar winds. However, the rates of both processes remain highly uncertain.
\citet{AlisterSeguel+2020} studied the effects of mass loss during mergers in SMS-forming clouds with masses and sizes similar to those in our simulations. They found that mergers resulted in a loss of 15--40\% of the stellar mass during the formation of a $10^5~M_\odot$ star. In their study, mergers were the dominant contributors to mass growth, accounting for 50--60\% of the total stellar mass. In contrast, our simulations indicate that mergers contribute a smaller fraction of the mass. Assuming the mass loss scales with the merger contribution, the expected mass loss in our calculations would be at most 20\%.
\citet{Nakauchi+2020} investigated pulsational mass loss rates during the main sequence and red giant phases for very massive stars with $M_* \lesssim 3000~M_\odot$, using linear stability analysis. They concluded that pulsational mass loss rates are $\lesssim 10^{-3}~M_\odot\mathrm{yr}^{-1}$. This suggests that pulsational mass loss is negligible during SMS formation, as the mass growth through accretion is orders of magnitude higher. However, once accretion ceases, pulsational mass loss could significantly reduce the mass of very massive stars ($M_* \sim 1000~M_\odot$).
In our simulations, stars with masses around $1000~M_\odot$ form at [Z/H]$=-2$. During their later red giant phase, line-driven winds could contribute to mass loss, reducing their mass by approximately 30\%. While this effect has a relatively minor impact on SMS formation, it could significantly affect the final masses of stars in this range.

We have demonstrated a remarkable transition in the nature of the final stellar systems, shifting from supermassive stellar systems to globular cluster-like systems at [Z/H]$=-2$.
At this metallicity, a stellar cluster forms, dominated by central VMSs with $M_* \sim 1000~M_\odot$. Once these central stars reach the end of their lives, the remaining cluster exhibits a stellar surface density of $10^3~M_\odot~\mathrm{pc^{-2}}$ and a half-mass radius of $\sim 4~$pc, excluding the massive stellar component ($M_* > 100~M_\odot$). These characteristics are similar to those of typical young massive clusters or globular clusters observed in the local universe.
Interestingly, the metallicity threshold we identify closely matches the floor of [Z/H]$\sim -2.5$ observed for globular clusters in the Milky Way \citep{Puzia+2005, Beasley+2019}, providing a compelling explanation for their origins.
However, the surface densities of these clusters are about an order of magnitude lower than those observed in strongly lensed high-redshift clusters, such as the Sunrise Arc at $z\sim 6$ \citep{Vanzella+2023} and Cosmic Gems at $z\sim 10$ \citep{Adamo+2024}, which exhibit densities of $10^4$--$10^6~M_\odot\mathrm{pc^{-2}}$.
Including the contribution of massive stars ($M_* > 100~M_\odot$), which dominate both UV photon emissivity and stellar mass during their lifetimes, significantly enhances the total stellar mass and surface density. The total stellar mass, largely contributed by these massive stars, reaches approximately $2 \times 10^5~M_\odot$, with a UV emissivity per unit mass far exceeding that of lower-mass stars. Observing these clusters while the central massive stars are still alive would reveal a dramatic increase in UV photon emissivity, by at least an order of magnitude, bringing the surface densities closer to those observed in strongly lensed high-redshift clusters.

The formation of very massive stars in star clusters has also been reported in \citet{Fujii+2024}, who demonstrated that stars with masses around $\sim 1000~M_\odot$ can form in low-metallicity environments of $Z = 0.02$ and $0.1~Z_\odot$. Their study showed that such massive stars could grow purely through stellar collisions, starting from an initial cloud profile similar to the one used in our simulations. Our findings complement this result by suggesting that gas accretion also plays a significant role in stellar mass growth. Specifically, gas accretion can increase the final stellar mass by an additional 50\% compared to growth achieved solely through mergers (Fig.~\ref{fig::mass_by_mergers}). This indicates the importance of the accretion as well as mergers when investigating the formation of massive stars in low-metallicity environments, as both processes together can significantly influence the resultant stellar mass and the evolution of the surrounding star cluster.

The presence of central massive stars offers a natural explanation for the chemical abundance patterns commonly observed in globular clusters, such as the anti-correlations between C-N and Na-Al abundances \citep{Carretta+2009, Carretta+2010}. Material processed within stars with $M_* \gtrsim 1000~M_\odot$ is expelled through stellar winds, enriching the surrounding environment with nucleosynthetic products. This process results in the observed anti-correlation patterns, accompanied by low helium abundances, as suggested by previous studies \citep{Renzini+2015, Bastian+2018, Gieles+2018}. For even more massive stars, with $M_* \gtrsim 5000~M_\odot$, additional signatures such as the Mg-Al anti-correlation are expected, reflecting the unique nucleosynthetic processes within such stars \citep{Denissenkov+2014}.
Moreover, the significant nitrogen enrichment from these massive stars may account for the abundance patterns observed in high-redshift nitrogen-rich galaxies recently identified by JWST \citep{Isobe+2023, Charbonnel+2023}. These findings highlights the importance of central massive stars in shaping the chemical abundance patterns observed in globular clusters and providing a natural explanation for the peculiar signatures seen in the early universe. Such systems, as we have demonstrated, naturally form in environments where gas enriched to [Z/H]$\sim -2$ is exposed to strong FUV radiation fields. This scenario offers a unified framework for understanding both the origins of globular clusters and the chemical enrichment processes of high-redshift galaxies, bridging the gap between local and early-universe observations.

%%%%%%%%%%%%%%%%%%%%%%%%
%%%%%%%%%%%%%%%%%%%%%%%%
\section*{Acknowledgements}
We thank Bastian Reinoso, Mark Gieles, Collin Charbonnelle, Raffaella Schneider, Rosa Variante, Gen Chiaki, Lars Hernquist, and Chiaki Kobayashi for fruitful discussions and comments.
This research was supported in part by
the Grants-in-Aid for Basic Research by the Ministry of Education, Science and Culture of Japan (KO:22H00149) and by grant NSF PHY-2309135 to the Kavli Institute for Theoretical Physics (KITP).
We conduct numerical simulations on XC50 \texttt{ATERUI II} at the Center for Computational Astrophysics (CfCA) of the National Astronomical Observatory of Japan and XC40, through the courtesy of Prof. E. Kokubo.
We also carry out calculations on XC40 at YITP, Kyoto University.
We use the SPH visualization tool SPLASH \citep{SPLASH} in Figs.~\ref{fig::snap_overall_density}, \ref{fig::snap_overall_Tgas}, \ref{fig::snap_evo}, \ref{fig::snap_overall_tMyr}, \ref{fig::snap_M-4} and \ref{fig::snap_M-2}.

\section*{DATA AVAILABILITY}
The data underlying this article will be shared on reasonable request to the corresponding author.

\bibliographystyle{mnras}
\bibliography{ms}

\appendix

\section{Resolution effect} \label{sec::resolution}
\begin{figure*}
\centering
\includegraphics[width=0.95\textwidth]{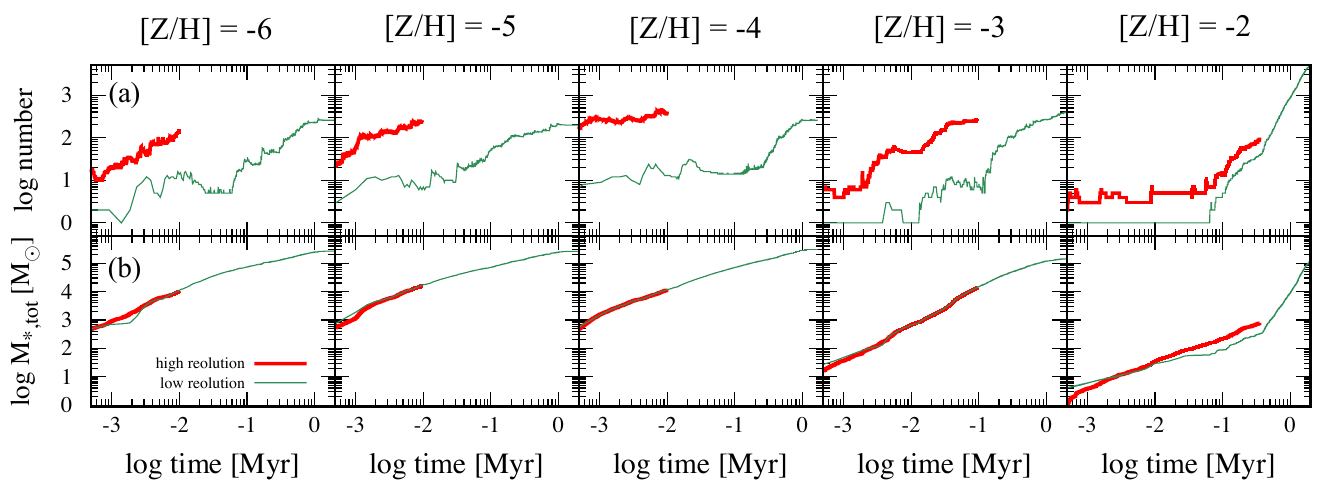}
\caption{
(a) Time evolution of the number of stars for high-resolution runs (red lines) and low-resolution runs (green lines) across different metallicity cases. The higher-resolution runs capture a significantly larger population of low-mass stars ($M_*\lesssim 10~M_\odot$), which are underresolved in the lower-resolution simulations.
(b) Time evolution of the total stellar mass for the same simulations. Despite the differences in resolving low-mass stars, the total stellar mass evolves similarly in both resolutions, indicating that the formation and growth of massive stars dominate the mass budget.
}
\label{fig::resolution}
\end{figure*}

Fig.~\ref{fig::resolution} quantitatively illustrates the impact of numerical resolution on the resulting mass spectra. The time evolution of the number of stars (panel a) and the total stellar mass (panel b) is compared for lower-resolution (green lines) and higher-resolution runs (red lines). Panel (a) shows that lower-resolution runs underestimate the number of stars, as they fail to resolve the population of low-mass stars with $M_*\lesssim 10~M_\odot$, which are captured in the higher-resolution runs (see Figs.\ref{fig::mass_spectra_t2Myr} and ~\ref{fig::mass_spectra}). In contrast, panel (b) demonstrates that the total stellar mass evolves similarly in both cases, indicating that massive stars dominate the mass budget. This consistency suggests that the assembly and growth of supermassive stars can be reliably reproduced even in simulations with lower resolution.

% Don't change these lines
\bsp	% typesetting comment
\label{lastpage}
\end{document}